\documentclass[%
 reprint,
 amsmath,amssymb,
 aps,
]{revtex4-2}

\usepackage{graphicx}% Include figure files
\usepackage{dcolumn}% Align table columns on decimal point
\usepackage{bm}% bold math
\begin{document}

\preprint{APS/123-QED}

\title{Empirical evidence for a jamming transition in urban traffic}

\author{Erwan Taillanter}
\email{erwan.taillanter@ipht.fr}
\affiliation{Institut de Physique Th\'{e}orique, CEA, CNRS-URA 2306, F-91191, 
Gif-sur-Yvette, France}

\author{Marc Barthelemy}
\email{marc.barthelemy@ipht.fr}
\affiliation{Institut de Physique Th\'{e}orique, CEA, CNRS-URA 2306, F-91191, 
Gif-sur-Yvette, France}
\affiliation{CAMS (CNRS/EHESS) 54 Avenue de Raspail, 75006 Paris, France}

\date{\today}% It is always \today, today,

\begin{abstract}
  
  Understanding the mechanisms leading to the formation and the propagation of traffic jams in large cities is of crucial importance for urban planning and traffic management. Many studies have already considered the emergence of traffic jams from the point of view of phase transitions, but mostly in simple geometries such as highways for example, or in the framework of percolation where an external parameter is driving the transition. More generally, empirical evidence and characterization for a congestion transition in complex road networks is scarce, and here we use traffic measures for Paris (France) during the period 2014-2018 for testing the existence of a jamming transition at the urban level. In particular, we show that the correlation function of delays due to congestion is a power law (with exponent $\eta\approx 0.4$) combined with an exponential cut-off $\xi$. This correlation length $\xi$ is shown to diverge during rush hours, pointing to a jamming transition in urban traffic. We also discuss the spatial structure of congestion and identify a core of congested links that participate in most traffic jams and whose structure is specific during rush hours. Finally, we show that the spatial structure of congestion is consistent with a reaction-diffusion picture proposed previously. 

\end{abstract}

\maketitle

%\tableofcontents

\section{Introduction}

%Over half of the worldwide population lives in cities. Those areas are often the main drivers of a country's economy, as they concentrate a high density of industries, universities and people in reachable distance from one another. However, the transportation of large numbers of commuters is a challenge for all large cities. Road users experience congestion and delays almost on a daily basis, resulting in the loss of valuable time, stress for the commuters or increased emissions with consequences for the environment and the population's health. %il manque des refs pour ça

The study of traffic and specifically its congestion has been a topic of interest for almost as long as individual transportation exists. The first works on the subject dealt with traffic on highways, applying various methods from fluid mechanics \cite{hydro model} to cellular automata \cite{cellular automaton}. Concepts like the carrying capacity of a highway, the formation and progression of congestion on a highway, or the impact of multilane roads with traffic travelling at different speed on the capacity are well understood and described in the literature \cite{Traffic theory on highways, Nagel Paczuski, Blandin theory for phase transition}. Sugiyama et al. \cite{Sugiyama et al. experiment} and Tadaki et al. \cite{Tadaki et al. experiment, Tadaki et al. correlation} have conducted experiments on one-lane roads confirming that the switch from free-flow to congestion is brutal and analog to a phase transition, while in \cite{Chacoma et al. traffic lights} it has been shown that traffic lights also induce a phase transition of the state of traffic on a single lane road. In \cite{Cuesta,Biham Middleton} models have been proposed for the formation of congestion on a regular 2D lattice, displaying a first-order phase transition between free-flow and jammed state.

In contrast, in more complex networks of roads, the mechanisms of congestion remain poorly understood. Urban road networks are strongly connected systems and scaling from single lane roads to a non-regular two-dimensional network is not an easy task. Agent-based simulations try to simulate the flow of cars on the network, based on the `microscopic' interaction between cars on the network, and between cars and the network itself (e.g. traffic lights). This type of of simulations requires many assumptions and parameters and are difficult to validate. More recent studies analyze the urban traffic at a macroscopic level, focusing on the description of the emergence of large-scale congestion as a phase transition. Indeed, it is known that the jamming of a node of the network which happens when the demand exceeds its capacity, can have a macroscopic impact, as shown by Echenique et al. \cite{Echenique} in the context of Internet protocols. However, the exact response of the network depends strongly on the network itself and on the way the information propagates on the network. Understanding the origin and the propagation of congestion on urban road networks is therefore the aim of most studies on the subject.

At a theoretical level, studies in \cite{Lampo et al.,Carmona et al.} have proposed arguments to understand the emergence of congestion on the network, based on its topology. In \cite{Geroliminis diffusion}, based on previous work \cite{Jiang Havlin} Bellocchi and Geroliminis 
proposed an application of reaction-diffusion equations to the case of traffic congestion, where the congestion tends to spread from a link to its neighbors. At the other end of the spectrum, Saberi et al. \cite{Saberi Gonzalez epidemio} proposed a model inspired by epidemiological studies to describe the evolution of the number of congested links, regardless of the actual structure of the network.

At a more empirical level, a lot of attention has been given to the study of congestion in the framework of percolation \cite{Stanley percolation} where the main idea is to look at the structure of the set of links at a certain level of congestion (which depends then on the hour of the day).
For each date, the authors of \cite{Stanley percolation} divide the network into functional and congested roads, based on a velocity factor $v^*$. This leads to clusters of roads functioning with a speed $v>v^*$, separated by congested roads with speeds $v<v^*$. By varying the threshold $v^*$, they observe a percolation transition (that depends on the hour of the day) with a breakdown of a giant functional cluster into several small clusters. The value of the percolation threshold $v^*_c$ can then be seen as a measure of the state of the network at that date, as it measures effectively the maximal velocity one can travel over the main part of the network (described by the giant component). Using a similar approach, Zhang et al. \cite{Zhang clusters encombrement} studied the size distribution and recovery times of clusters of congested links, and in particular the power-law dependence of the recovery time on the cluster size. Furthermore, Zeng et al. \cite{Zeng et al.} have shown that the randomness in the traffic demand and the emergence of local congestion means that the transition between the giant functional cluster and several small clusters is not monotonous, but displays switches between multiple metastable states.
Cogani et al. \cite{Cogani first idea, Cogani real article} studied the spatial distribution of the clusters and found in agreement with \cite{Zhang clusters encombrement} that, once formed, the clusters are stable in time, from day to day, indicating that their existence is determined by the traffic patterns (origins and destinations of the road users) and by the network itself. This approach opened new perspectives to analyse a road network, allowing in particular to identify the bottlenecks of the network, responsible for the collapse of the giant component into smaller clusters \cite{Hamedmoghadam bottlenecks from clusters}.

In these studies, the percolation threshold is a convenient way to characterize the level of congestion in the network but doesn't provide evidence that there is some kind of jamming transition as it is observed in simpler systems \cite{Kadanoff:2000}. In particular, a common signature of phase transitions is the divergence of the some correlation length near criticality \cite{Kadanoff:2000}). Guo et al. \cite{Guo et al.} identified the most influential roads of the network based on their correlation with the traffic on other roads later in time, and found that roads could have a significant impact on traffic up to $1km$ away (in the case of Beijing), indicating strong correlations in the network. In \cite{Petri et al.}, it has been shown that London could be partitioned into clusters based on the intrinsic measure of the correlation between roads. Roads with strongly correlated measures of traffic were associated into clusters which sometimes spanned the entire network. More importantly, the correlation function for traffic fluctuations over the whole network decays as a power-law during rush hours, a typical signature of a critical phenomenon.

Here, we focus on the empirical characterization of a transition in road networks but using the delay experienced by users on each road as the most relevant quantity for defining the congestion level. We compute the correlation function for the network at each hour and identify a typical power law with an exponential cut-off. From this result, we identify the correlation length and we find that it diverges during rush hours pointing to the existence of a phase transition. We also discuss the peculiar behavior and its impact on the network of the outer ring road during rush hours, displaying much higher correlation levels than the rest of the network. We then analyze the spatial distribution of congested roads. As in \cite{Zhang clusters encombrement}, we build the cluster of congested links, which percolates the network during rush hours.
We identify roads appearing in this congested cluster during the majority of days and roads less prone to congestion and show that these results can be interpreted with a simple diffusion based model in agreement with previous studies \cite{Geroliminis diffusion, Cogani real article, Jiang Havlin}.

\section{Data, critical quantities and context}

\subsection{Dataset}

We use traffic measurements provided by the city of Paris \cite{DataParis} for the years 2014-2018. The most important roads of the Parisian network have been equipped with sensors in the early 2010s which are of two different types: some measure the traffic, i.e. the number of cars passing during a certain time-span, while the others measure the occupancy rate, i.e. the ratio between the time where a car is above the sensor and the total time. Note that here a `road' doesn't correspond to the traditional definition of a street with a name. Indeed, each street is subdivided into different segments at each intersection. The presence of this intersection affects the traffic and leads to different behaviors on consecutive segments. In the absence of intersections however, the traffic properties will be constant on the segment and doesn't need to be further subdivided. We thus consider the graph where the links are the road segments that connect the nodes which represent intersections. For highways for example, a link would then connect an entry and the next exit.

Not all road segments are equipped with both types of sensors and their exact number varies from year to year. Roughly, the number of links equipped with both traffic and occupancy sensors is about $10\%$ (around 1,000, out of approximately 10,000 for the whole network). All the results that we give here and which involve a fraction of the links of the network will implicitly be a fraction of the number of links for which a `dual' measure is possible.

The dataset made publicly available spans from 2014 to 2019 and has a 1 hour time step. This means for example that the traffic on link $l_i$ at 9am corresponds to the total number of vehicles that travelled on $l_i$ from 8am to 9am. Similarly, the occupancy rate is the fraction of the time during which a car was present on the sensor from 8am to 9am. All our results will follow the same logic, where a result for a given hour $h$ refers to the average from $h-1$ to $h$.

We will focus in this work on weekdays when the congestion levels are highest. We also removed August from our data set, since congestion levels are significantly lower during this month (when most parisians are in vacation).

\subsection{Relevant quantities}

Raw data provided by the city of Paris are not directly of use for our study and have to be transformed. More precisely, the data provided by the city of Paris is under the form of two quantities. First the occupancy rate $o_i(t)$ of a road segment $i$ at time $t$ - a rather uncommon quantity for analyzing road traffic - corresponds to the presence of cars on the sensor loop measured as the percentage of every hour. For instance, an occupancy rate of $25\%$ means that cars have been detected on the loop during $15$ minutes. It is in general more common to use the density $k_i(t)$ of cars on a link $i$ at time $t$ defined as the number of cars per unit length. Under the assumption that sensors are not too close to traffic lights, and that all cars have roughly the same length, these two quantities are however proportional. Indeed, if we denote by $L_c$ and $L_s$ the typical length of a car and a sensor respectively (with typical values $5$ and $1$ meters), the density of cars $k_i(t)$ and the occupancy rate $o_i(t)$ (for a link $l_i$ at date $t$) are related by
\begin{align}
    k_i(t) = \frac{o_i(t)}{L_c+L_d}
\end{align}

The other quantity provided in the dataset is the traffic flow $q_i(t)$ that measures the number of vehicles that have passed a given sensor during a fixed time interval (which is here $1$ hour). Knowing both the traffic flow $q_i(t)$ and the density $k_i(t)$ of cars on the road segment $l_i$ at date $t$ allows then to estimate the travel speed $u_i(t)$ on this link (averaged over the time step)
\begin{align}
    u_i(t) = \frac{q_i(t)}{k_i(t)}
\end{align}
A well-known fact in traffic theory - mostly for simple cases such as highways \cite{Traffic theory on highways} - is the relation between the car density and the flow
which is described by the so-called fundamental diagram \cite{Traffic theory on highways}. When the car density increases the flow grows, reaches a maximum and then decreases when congestion effects become important (examples for road segments in Paris of this Fundamental Diagram are provided in the Appendix, see Fig. SM 1). At low density,  the traffic flow is proportional to the density and this linear relation allows to evaluate the maximal speed $u_i^{max}$ (i.e. the free flow velocity) for each road segment. The free flow velocity varies strongly from a link to another and we find that for Paris, it is log-normally distributed among the links. It is natural to normalize the velocity on each link by it free flow value and to define $v_i(t) = \frac{u_i(t)}{u_i^{max}}$, which is the standard quantity used for instance in \cite{Stanley percolation}. Here, we want to define a measure of the macroscopic state of the network, i.e. an aggregation of the states of the links, and the natural quantity is then the relative increase of travel time (or relative delay) $d_i(t)$ on each link. For a link of length $L_i$, it is given by
\begin{align}
  \nonumber
  d_i(t) &= \frac{L_i/u_i(t)-L_i/u_i^{max}}{L_i/u_i^{max}}\\
  &= \frac{v_i(t)-1}{v_i(t)}
\end{align}
Using this delay, we can compute the average delay over the whole network
\begin{align}
\label{eq network}
    d_{network}(t) = \frac{\sum_{l_i} d_i(t)}{N_{links}}
\end{align}
or other quantities such as the time lost on an average commute. Another important quantity is the length of the links experiencing delays which, for a same average delays, will vary between a high local congestion or a mild but generalized congestion. To do so, we define a congested link as a link on which the travel time has been increased by more than $10\%$.

\subsection{The Parisian network}
%\paragraph{Geography and commuting patterns}

The data concerns here Central Paris (`Paris Intra-muros') which is surrounded by a ring road (the `Boulevard P\'eriph\'erique') which plays an important role for commuters going from one part of the suburb to another one. Central Paris is roughly circular with a diameter of order $10kms$. For most inhabitants of the urban area that comprises central Paris, it is common to live and work in different places. The average commute to work is of order $6.6 kms$ for Parisians and of order $14.6 kms$ for individuals living in the suburb \cite{Commudata}. This means that the typical trip spans a significant part of the network.

Another interesting feature of the network is the impact of the Seine river crossing the city. As in most cities with a river, there are roads along it on each bank. In addition in the case of Paris until 2016, there was an urban highway (called the `Voie sur berge Georges Pompidou') going from West to East through Paris, while there were regular roads (smaller and with traffic lights) for the direction from East to West. The mayor of the city of Paris decided in 2016 to close this West-East highway in order to reduce air pollution and to give some space back to pedestrians. As a result, the dataset studied here effectively contains two quite different networks and traffic structures depending on the period of focus. Our results will be displayed comparing years 2014 to 2018, with 2016 being split into its beginning and its end (the closure occurred during the fall of 2016).

\begin{figure}[h!]
    \includegraphics[width=0.5\textwidth]{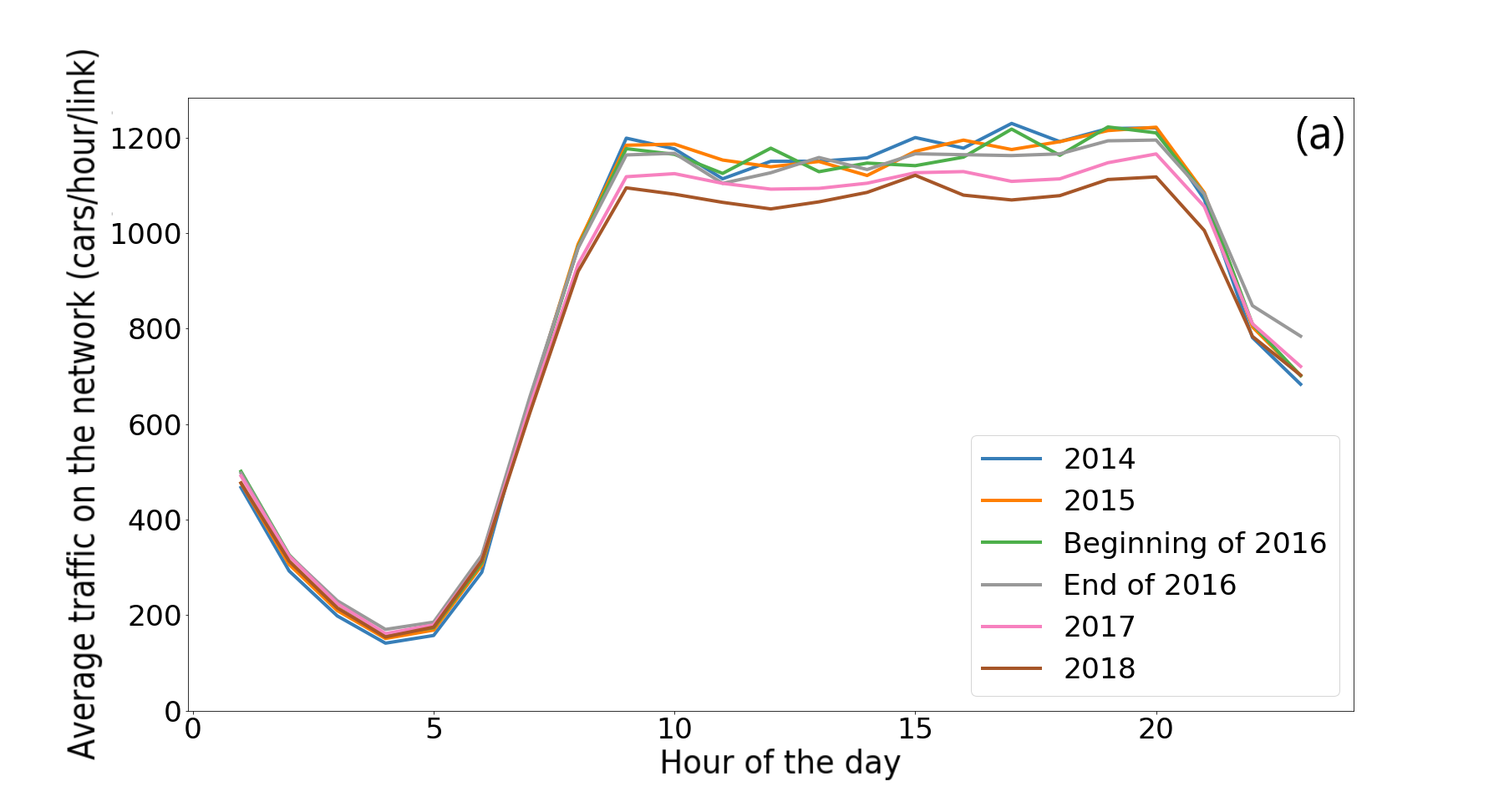}
    \includegraphics[width=0.5\textwidth]{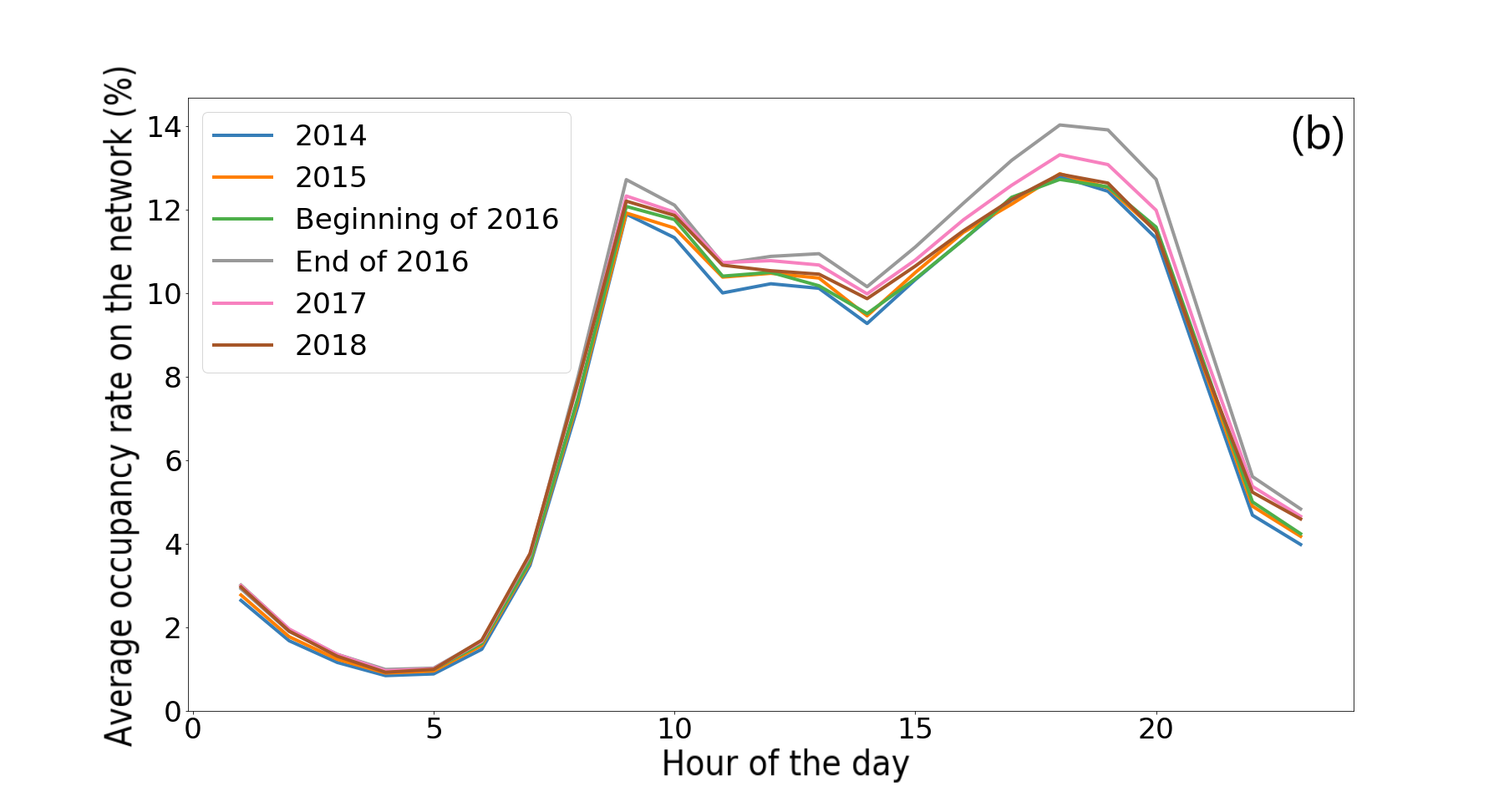}
    \includegraphics[width=0.5\textwidth]{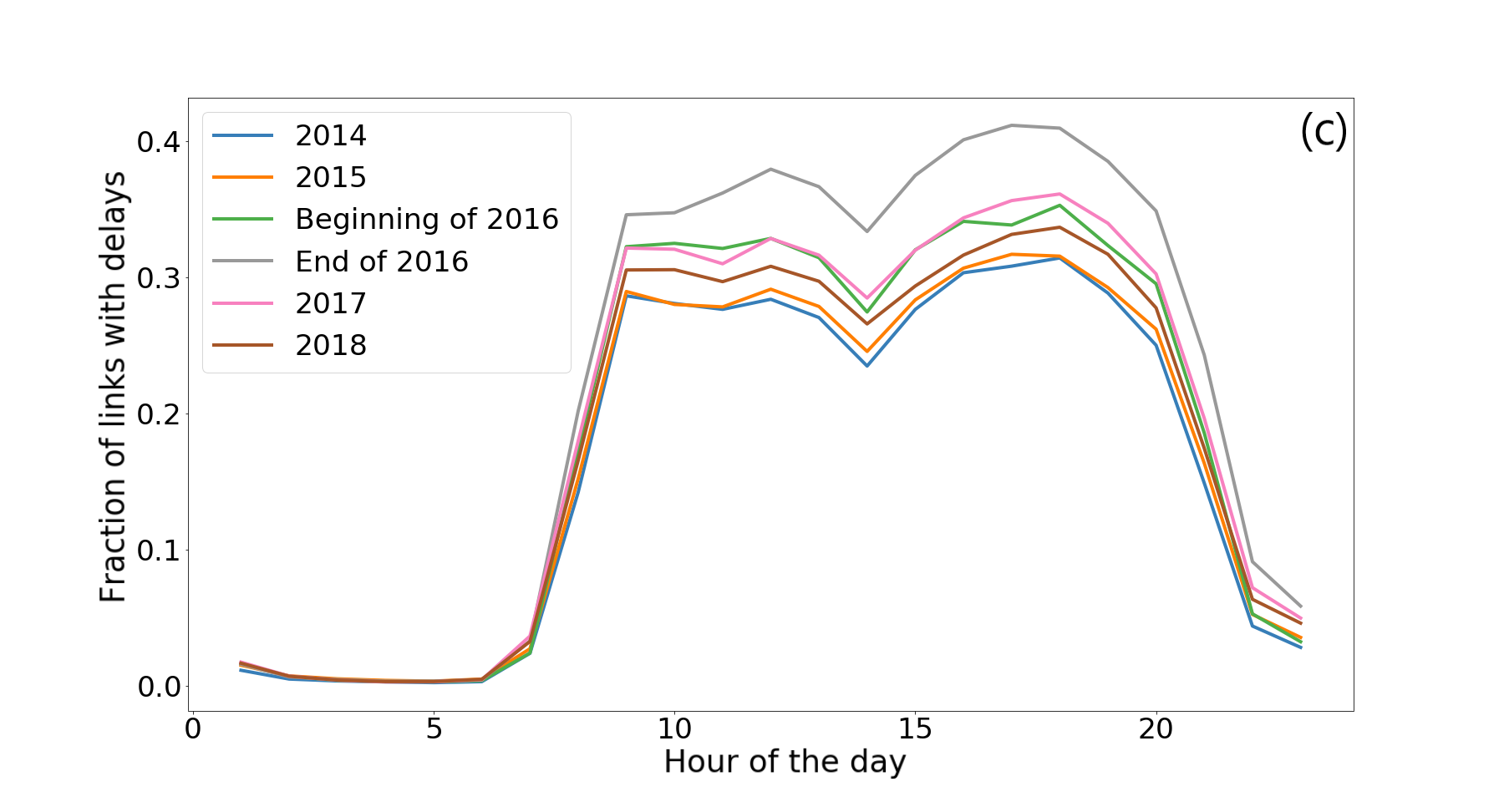}
    \caption{Daily evolution for years 2014 to 2018 of (a) the average traffic, (b) the average occupancy rate, and (c) the average fraction of links of the network experiencing delays. The traffic level is high during the whole day and rush hours can be observed on the occupancy rate and delays at about 8am and 7pm.}
    \label{Traffic measures Paris}
\end{figure}

%\paragraph{Traffic measures}
The focus of this paper is the formation of congestion and we will therefore consider working days and remove weekends and the month of August when the traffic demand is (much) lower. Unless specified otherwise, all the figures displaying average quantities are obtained by averaging over all working days in a year at a given hour.  For example, we show in Fig.~\ref{Traffic measures Paris} the daily evolution of relevant measures of traffic in Paris for the studied period. Interestingly, we note that the traffic level is high for most of the day without large differences between rush hours and mid-day. In contrast, the occupancy rate and the fraction of network with delays display more pronounced peaks during rush hours (in the morning around 8am and in the evening around 7pm). In addition, we note that even during rush hours there is about $30$ to $40\%$ of the total number of roads of the network that experiences significant delays, indicating a concentration of delays in a few `hotspots', in agreement with the results in \cite{Stanley percolation,Guo et al.}.

\section{Evidence for a transition}

The goal in this work is to find an intrinsic marker of a phase transition in the data and to show that congestion in a complex road network during rush hours can be viewed as a sort of jamming transition, going beyond one-dimensional cases. A typical marker of a phase transition in classical statistical physics is the divergence of the correlation length close to the critical value of the control parameter (see for instance \cite{Kadanoff:2000}). We will therefore analyze the correlation between the delays on all roads of the network, and identify the correlation length and its variation. We will show that it `diverges' (ie. it becomes comparable to the size of the network) during rush hours suggesting the existence of a transition. In addition, we analyze the impact of the ring road around Paris on spatial correlations of the delay. Note that this approach is similar to a previous work \cite{Petri et al.} with a few major differences that we will discuss in details in the following part.

\subsection{Correlation function}

As mentioned above, we compare roads based on the relative delay experienced by users. For each year and each hour, we thus have the delay $d_i(t)$ experienced on link $i$ and we introduce the quantity $T_i(t)$ which denotes the average travel time for that year and hour on this link. The relative delay is then given by $\tau_i(t)=d_i(t)/T_i(t)$ and indicates the importance of congestion on this link. We consider the correlation function of this quantity and measure it for the year $y$ and hour $t$
\begin{align}
  C_{i,j}(y,t) = \langle\tau_i(t)\tau_j(t)\rangle-\langle\tau_i(t)\rangle\langle\tau_j(t)\rangle
  \label{eq correlation}
  \end{align}
  The averages (denoted by the brackets $\langle\cdot\rangle$ are performed over all working days of a given year $y$ and at a given hour $t$. From this definition, we construct a distance dependent correlation function by sorting the pairs $i$, $j$ according to their distance. Since we are considering a phenomenon that takes place on a network, some caution is needed for defining the distance.  As discussed in \cite{Okabe}, the naive choice of the euclidean distance is not adequate as all physical processes discussed here occur on the network itself which somehow distorts distances. Instead, one should consider in general that each pair of roads is separated by the distance along the shortest path on the network connecting them. Here we will use the shortest travel time between roads (we use the path which, on average, is the shortest for a given year and hour). Petri et al. \cite{Petri et al.}, on the other hand, computed the correlation between traffic fluctuations in London (i.e. for each time step, the difference between measured traffic and expected traffic on each road). There are however some limits of using the values of traffic for measuring the state of the congestion state of the network. Indeed, if we assume that we have two links A and B on which the congestion level is growing in a correlated way. Depending on the fundamental diagrams of A and B (and on the initial state of congestion), one could have an increase of traffic on A and a decrease of traffic on B, despite both being more congested. As a result, the correlation of traffic fluctuations between A and B is not an ideal measure of the real correlation of the traffic state between A and B. The authors of \cite{Petri et al.} also used the euclidean distance to derive the correlation function and found a power-law decaying correlation function. Choosing the euclidean distance rather than the travel time implies that different types of roads are mixed together (see below of a discussion on this point).

 We found in our dataset that the correlation function displays a typical behavior of the form
\begin{align}
    C(r,y,t) = \frac{1}{r^\eta}\exp(-\frac{r}{\xi(y,t)})
\label{equation correlation length}
\end{align}
where $r$ is the distance between two roads and where $\xi(y,t)$ is the correlation length, depending on the hour of the day $t$ and calculated for each year $y$. As mentioned earlier, distances, including the correlation length, are converted in the time needed to travel them. The data and the corresponding fit are provided in the appendix (Fig. SM 2). As in \cite{Petri et al.}, we find that very short range correlations (i.e. between one link and its immediate neighbors) are strongly affected by the exact configuration of the network and extrinsic parameters such as the timing of traffic lights. For this reason, we performed our fits for distances greater than $60$ seconds of travel time, which corresponds roughly to the roads separated by more than just one traffic junction. Only a very small number of roads is distant of more than $3000$ seconds of travel time. Above this distance, the values of correlation are not statistically significant and we don't use them for the fits.

\begin{figure}[h!]
    \includegraphics[width=0.5\textwidth]{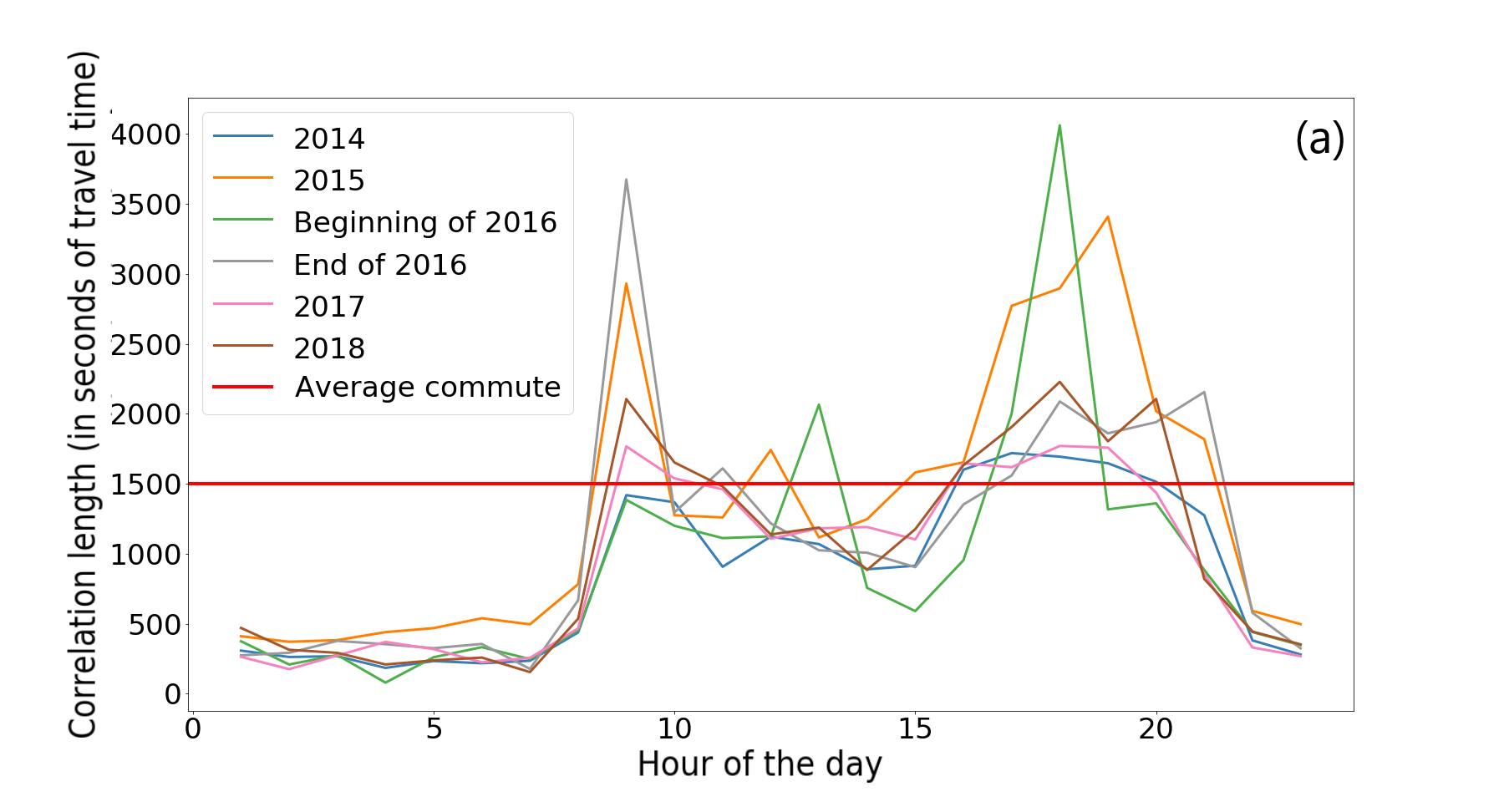}
    \includegraphics[width=0.5\textwidth]{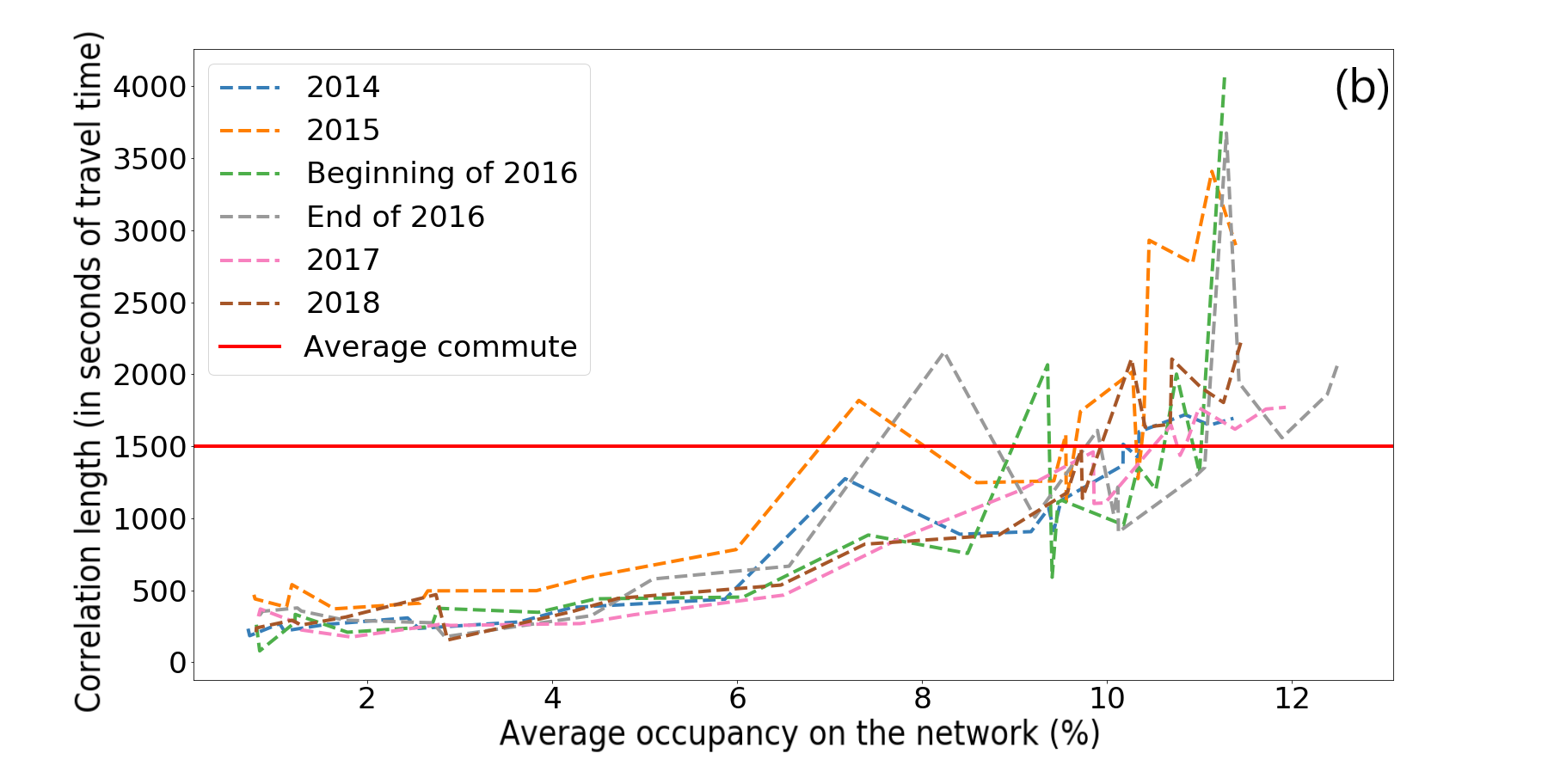}
    \caption{Correlation length $\xi$ for the delays on the links of the network, as function of time (a) and of the average occupancy rate (b). The horizontal line indicates the duration of a typical commute (1500 seconds), which geographically approximately corresponds to the size of the system}
    \label{Correlation length}
\end{figure}

This correlation function depends a priori on the year and the hour of the day. Performing the fits for all these cases, we are able to extract the correlation length and to examine its variation shown in Fig.~\ref{Correlation length}(a). First, we obtain consistently a value of $\eta\simeq 0.4$, which is higher than the value ($\approx 0.26$) found in \cite{Petri et al.}. Another major difference is of course the presence in our fit of an exponential cutoff for the correlation. In addition to the fact that we don't consider the same quantity for computing the correlation, the choice of euclidean distance in \cite{Petri et al.} might also affect the results. Indeed, two roads far away in space could be either separated by a highway and a short travel time apart (case (1)), or separated by the urban network and a long travel time (case (2)). We argue that these two cases translate in different correlations, with case (1) being more correlated than (2). In \cite{Petri et al.}, pairs of roads in both cases (1) and (2) are averaged over, resulting in larger values of correlation at long distances, which might explain why they don't observe an exponential cutoff in their results. The non-linear relation between travel-time and euclidean distance, provided for our network as an appendix (Fig. SM 4) might also explain in part the difference between the values of the exponent $\eta$.

We also observe a clear difference between night and day hours. During night hours, the correlation length is very small, typically of the order of the time needed to travel from a link to two or three links further:  delays appearing on a link during the night propagate to the vicinity of this link only. In sharp contrast, we observe during day hours a dramatic increase of the correlation length to values close to $1,500$ seconds and during rush hours, the correlation length displays a peak with values well above $1,500$ seconds. We recall here that the typical trip of a road user on the network is known to be around $1.500$ seconds long and having $\xi\approx 1,500s$ indicates a correlation at the scale of the whole system. This divergence 
is the sign of a transition that occurs during rush hours with congestion expanding over the whole network.
      
%\subsection{Control parameter and order parameter in the system}

These results therefore suggest the existence of a transition between a state where congestion is localized to another state where congestion occurs over the whole network during rush hours. The control parameter which varies with time is the traffic demand and it is then natural to consider the average density (or the average occupancy rate) on the network as the control parameter here. This leads us to investigate the relation between
the correlation length and the average occupancy rate on the network (see Fig.~\ref{Correlation length}(b)). As expected, the correlation length increases with the occupancy rate, and grows very rapidly for occupancy rates above $10\%$. However, the resolution of our data (one measure per hour), doesn't allow us to fit the observed function with enough precision, but this certainly constitutes an interesting point for further studies with more detailed data and could allow to determine the critical exponent $\nu$ for this transition.

\subsection{Distinctive role of ring roads}

In most large cities, there is a ring road in order to reduce traffic in the urban center by proposing an alternative route around the city. It is therefore important to understand its role in the spread of congestion. In the case of Paris, the ring road is called `Boulevard P\'eriph\'erique' and is encircling the central Paris area (and its length is about $35 kms$). We compute the correlation function for the links belonging to this ring road and the results are very different from those for the entire network. More precisely, the Boulevard P\'eriph\'erique displays two distinct regimes. During the night, as in the rest of the network, the correlation is small and local, but during the day the correlation is almost independent from the distance. It is maximal at short distances, but then remains large regardless of the distance. The Boulevard P\'eriph\'erique being an urban highway, we can apply standard methods of traffic theory to understand this result (see for example \cite{Traffic theory on highways}). In particular, it is well known that a traffic jam propagates on a highway like a wave with a speed given by
 \begin{align}
     u_w = \frac{q_d-q_u}{k_d-k_u}
 \end{align}
 where $q_{d(u)}$ and $k_{d(u)}$ are the downstream (upstream) flow and density. If we consider rush hours during which the Boulevard P\'eriph\'erique is congested both upstream and downstream, using the Fundamental Diagram (see Appendix, Fig. SM 1) we find that the congestion wave moves against the direction of traffic with a speed of order $50 km/h$. Considering the fact that this ring road is about $35$ kms long and that the peak for transport demand during the morning or the evening rush is over 2 hours long, it is clear that any unexpected incident on the Boulevard P\'eriph\'erique has the time to spread over the whole ring, explaining the high correlation values observed in our data. If the ring road is not congested, an incident will simply vanish locally and the correlation is small.
 
 We expect that this phenomenon happens for any large city with a highway-like ring road used above capacity during the day. This almost binary switch between small, short range correlations and large correlation independent from the distance on the ring road has an impact on the entire network. Indeed, roads that are on opposite sides of the city, but both close to the Boulevard P\'eriph\'erique will see their congestion level correlated, especially when the boulevard is strongly congested. In other words, the boundary conditions of the system change during the day and induces correlations between locations far apart from each other. During low demand hours, the P\'eriph\'erique acts as an open boundary (no correlation), while during rush hours, it can virtually be reduced to a single point. In other words, it acts during rush hours as a propagator of congestion at the boundary of the whole city.

 \section{Spatial structure of congestion}
 
 In this section, we investigate the spatial structure of congestion in Paris. In particular, we will focus on cluster of delayed roads (as done for example in \cite{Zhang clusters encombrement}) in order to analyze the spreading process of the congestion on the network.

\subsection{Giant component}

\begin{figure}[h!]
    \includegraphics[width=0.5\textwidth]{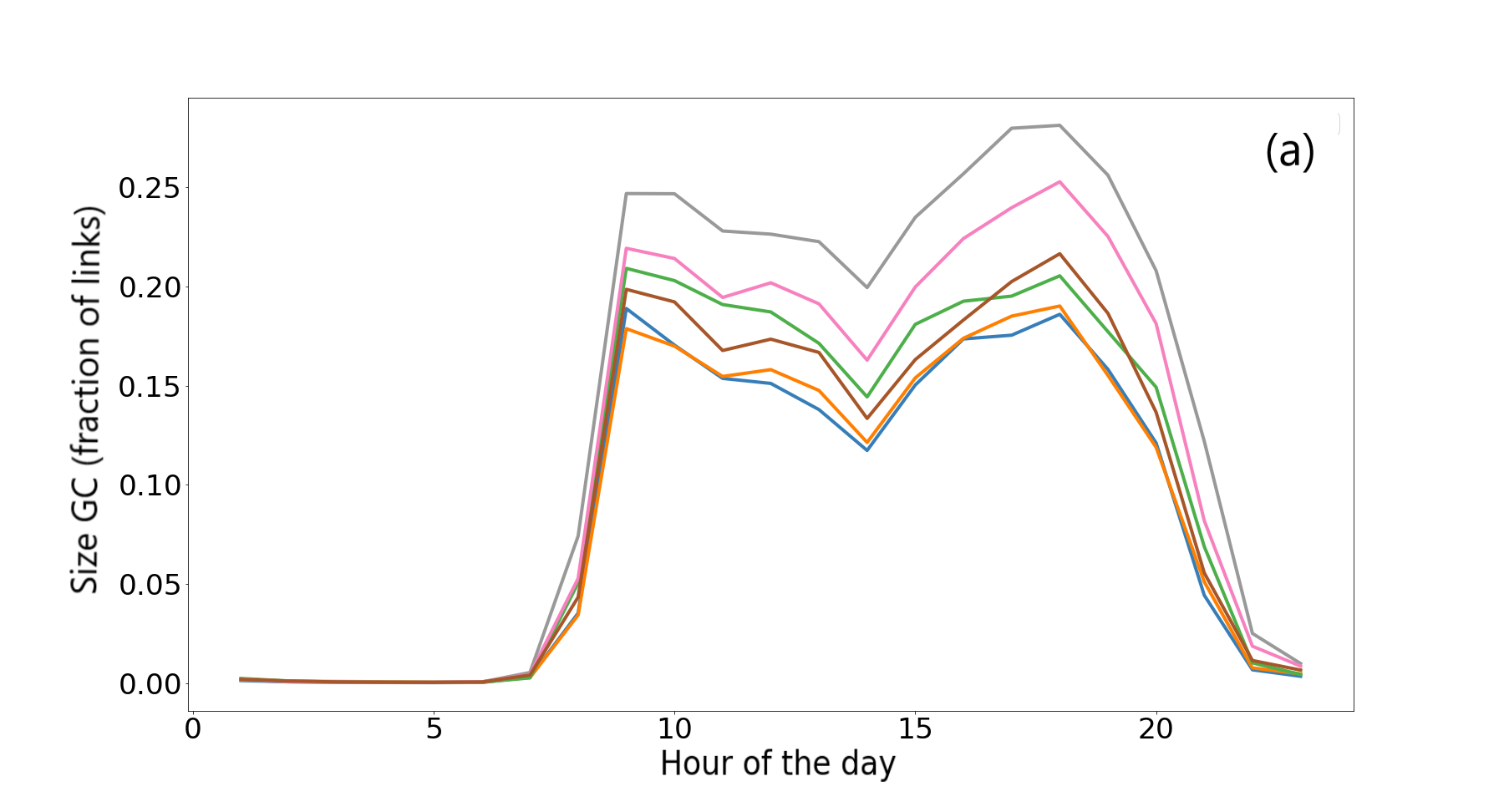}
    \includegraphics[width=0.5\textwidth]{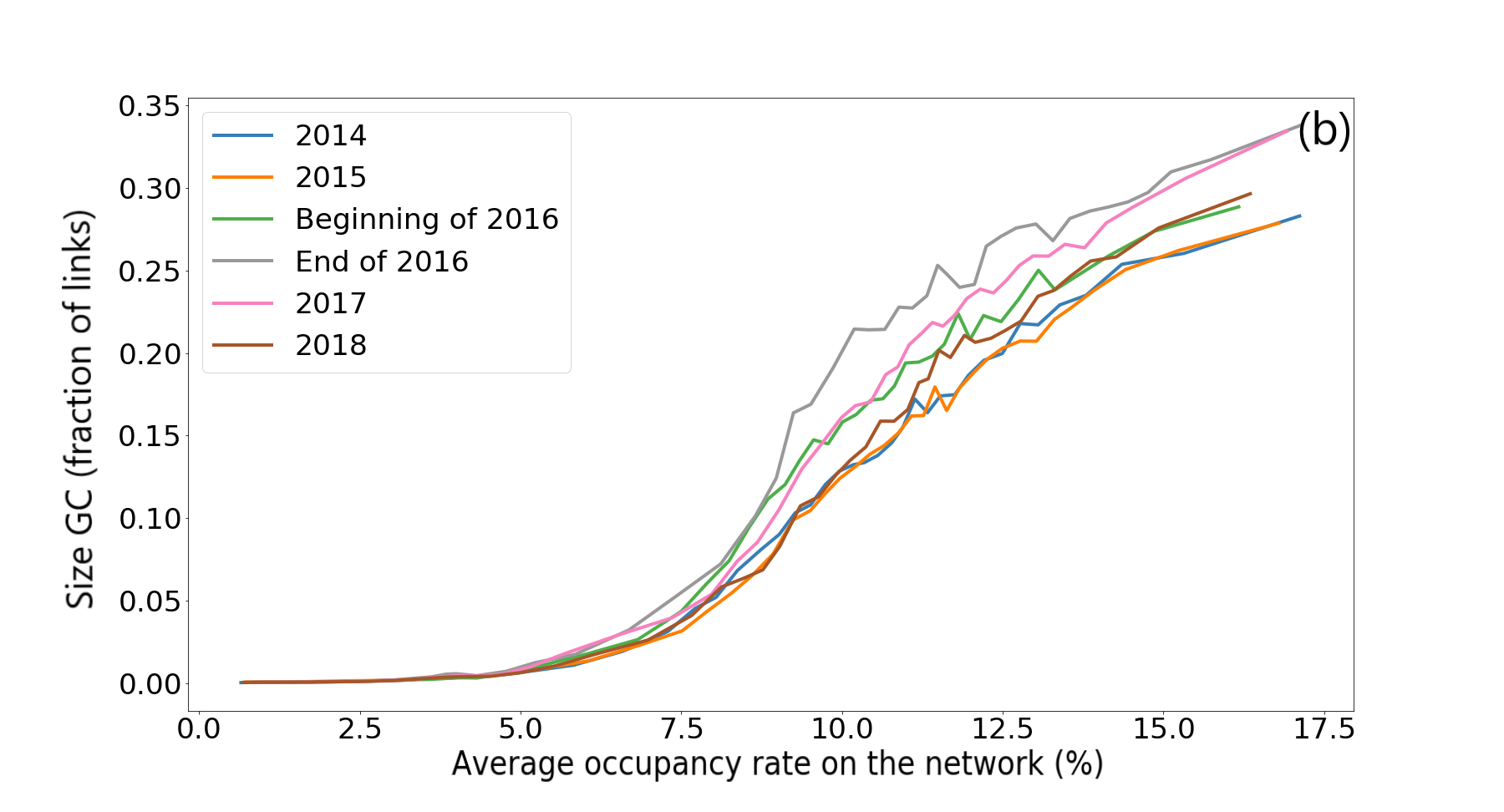}
    \caption{Average size of the Giant Component (fraction of links), depending on the hour of the day (a) and on the average occupancy rate (b). Averages performed over all working days of a given year at a given hour.}
    \label{size GC}
\end{figure}

For each hour, we define congested roads as those on which the travel time is increased by more than $10\%$ compared to the free flow situation. These roads form clusters of congestion on the network. Due to the limited spatial resolution of the data-set, we merged neighboring congested clusters based on the following rule: (i) if two clusters of congestion are separated by a path shorter than 350 meters (which corresponds to the typical length of a link) and (ii) if along this path, roads are not equipped with sensors (making it impossible to evaluate the delay), we consider that the two clusters are connected. For each date, we then look at the largest connected component of congested links. We find that the congestion percolates during rush hours and the dependence of the average giant component size on the average occupancy rate displays a typical shape of percolation in a finite size system, as displayed in figure \ref{size GC}(a).
Zeng et al. \cite{Zeng et al.} have shown that during rush hours the network oscillates between multiple meta-stable states: for a given fraction of congested links, they found that there could either be a giant functional cluster or several small functional clusters. We tried to reproduce these results for our case of the cluster of congestion rather than functional clusters. However, we did not find these meta-stable states and each measure of the size of the GC as a function of occupancy is close to the average relation displayed in Fig.~\ref{size GC}(b). The most likely explanation for this discrepancy is that our measures are averaged over an hour, which is a long period of time compared to the typical stability duration of the metastable states found in \cite{Zeng et al.}, which is of order $10$ minutes. This suggests that our measure thus only reflects the average of the systems oscillations between the two states.

\subsection{Congestion core}

\begin{figure}[h!]
    \includegraphics[width=\linewidth]{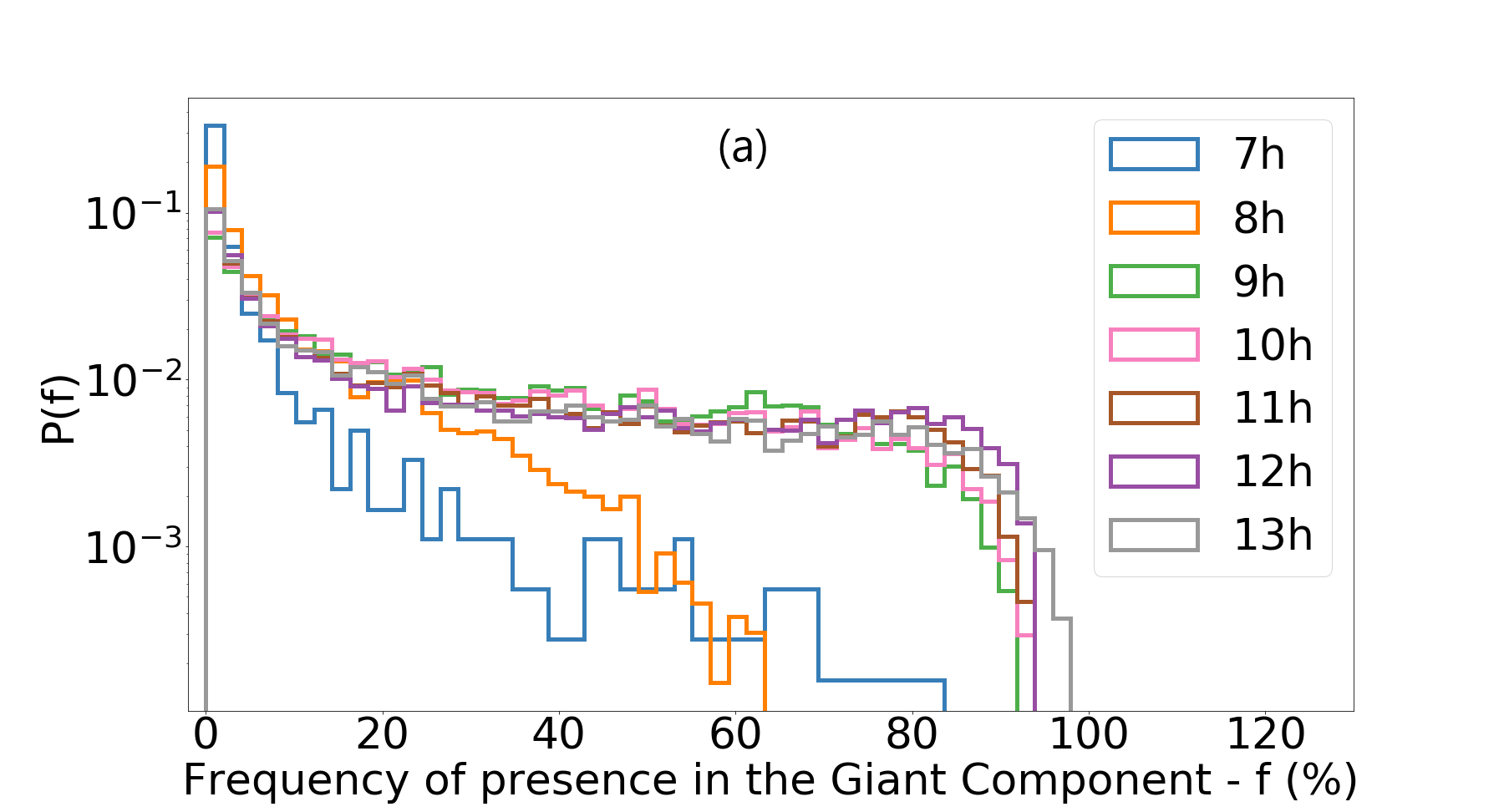}
    \includegraphics[width=\linewidth]{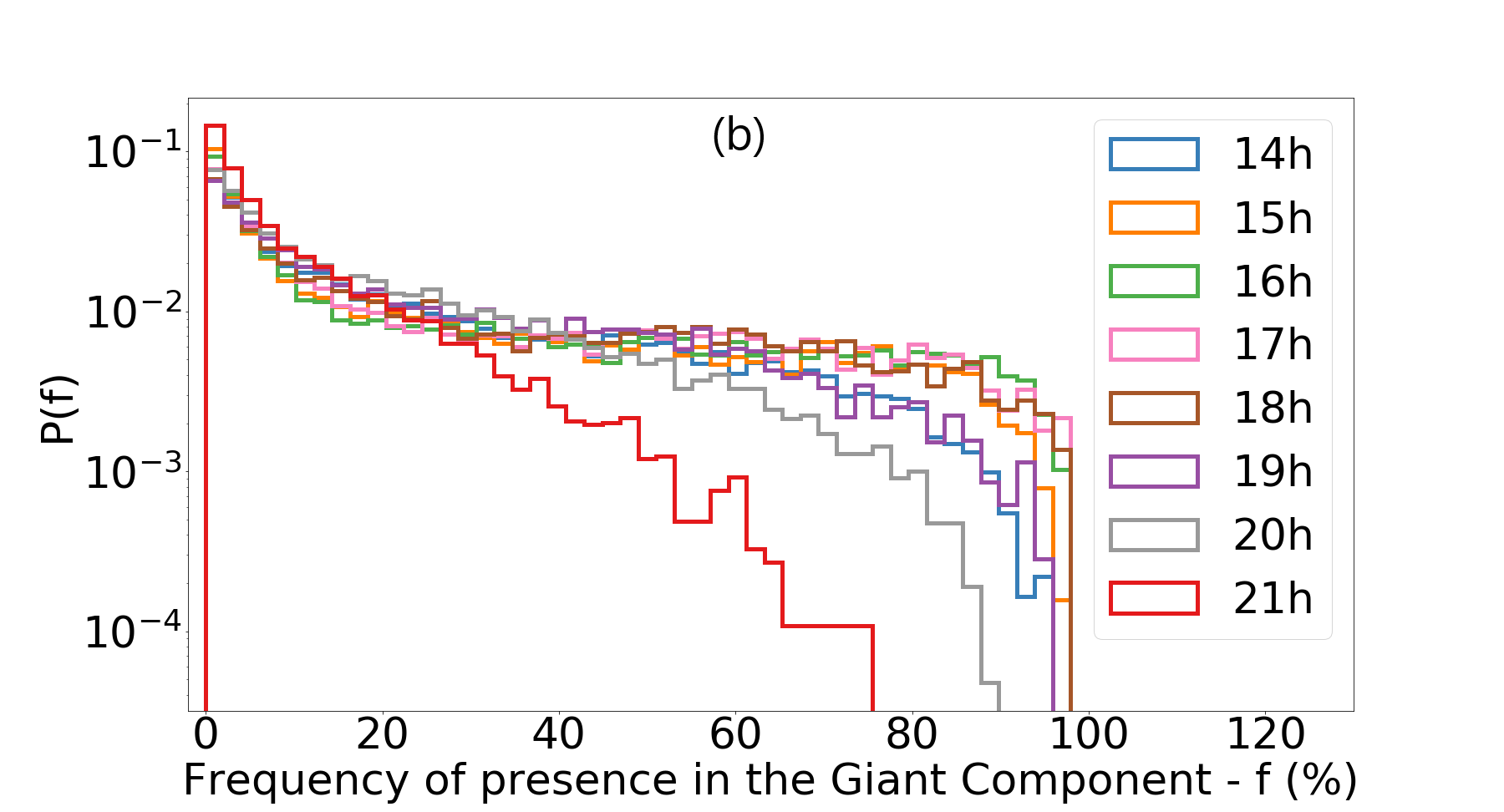}
    \caption{Distribution among all links of the frequency of presence of the link in the giant components of congestion of the year 2018 for morning (a) and afternoon (b) hours respectively. Note the stability between hours.}
    \label{frequency distribution}
\end{figure}

Once we have defined the giant component of congested links for each day and hour, it is natural to look at its stability in time. We thus identify each link that contributed to the giant component throughout the year and plot the distribution of its appearance frequency in the giant component (Fig.~\ref{frequency distribution}).
This result allows us to identify `core' links that are present in the giant component for most days and other links that appear only occasionally in the giant component. Examples of cores for different year and hours are provided in the Appendix (Figs. SM 5) and we can make some observations. First, there isn't a core of congested links during the night as few roads are congested and vary from a day to another. During rush hours, the core of congested links consists mostly of the Boulevard P\'eriph\'erique and other roads on the outer part of Paris, while during the other hours of the day, the core of congested links is smaller and concentrated in the center of Paris.

\begin{figure}[h!]
    \includegraphics[width=0.5\textwidth]{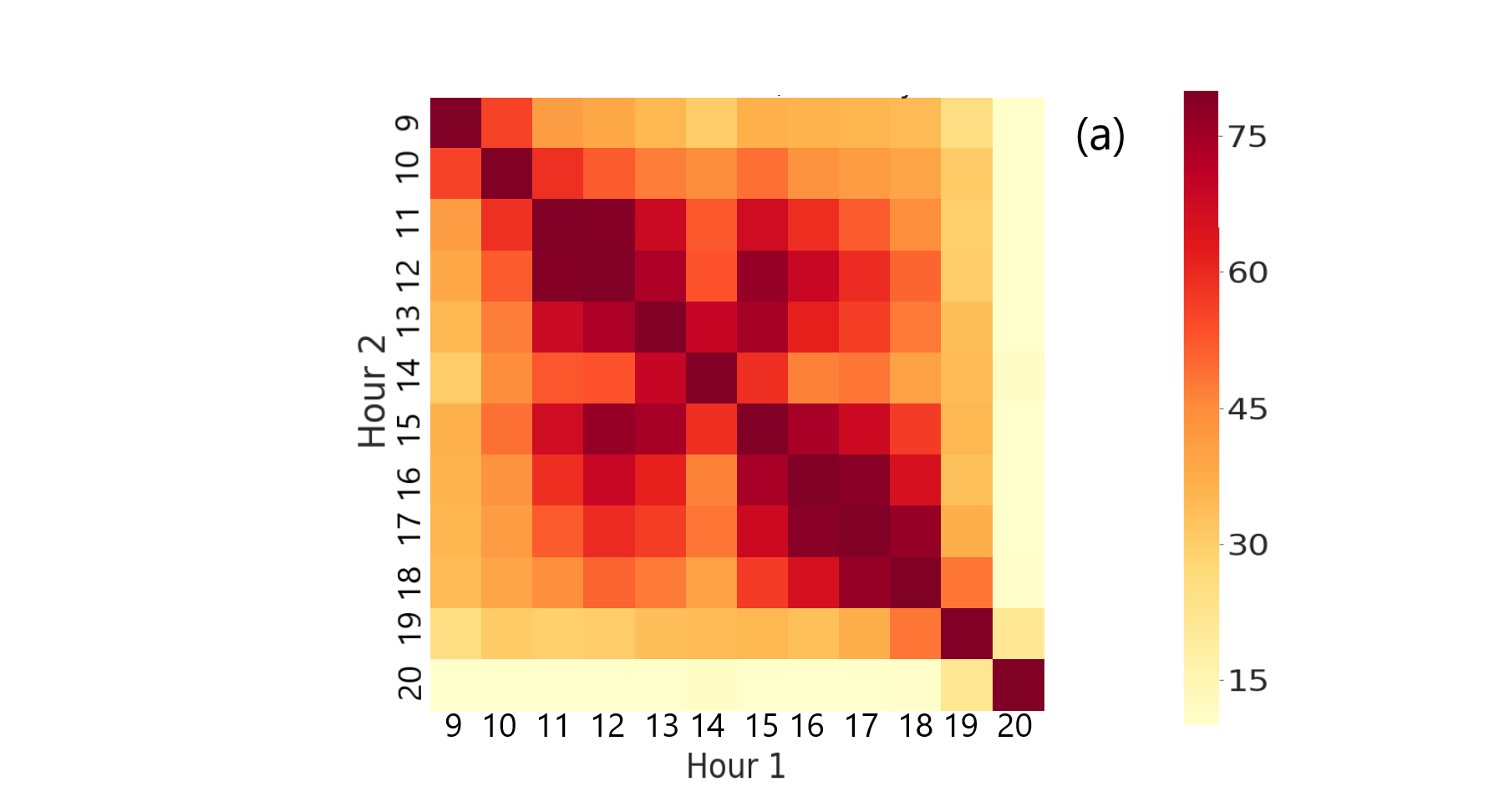}
    \includegraphics[width=0.5\textwidth]{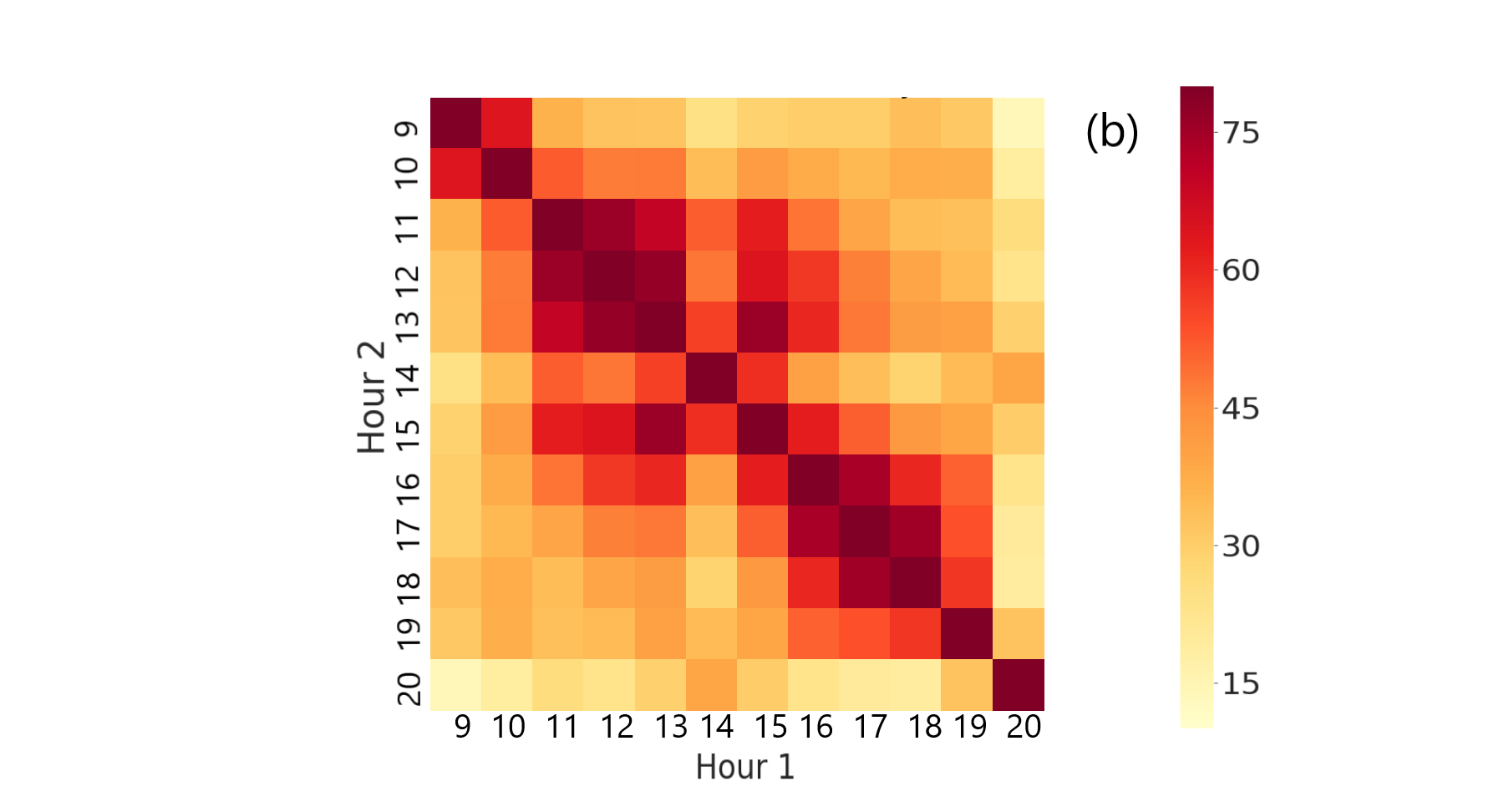}
    \caption{Overlap between the cores of congested links from hours to hour, for 2014 (a) and 2018 (b) respectively. One can clearly identify a continuity in the congested links during the morning and the evening rush, but only a small overlap between the two periods.}
    \label{Evolution noyaux}
\end{figure}

These results are consistent with the expected flows of commuters, coming from the suburbs to the center of Paris via the P\'eriph\'erique, but the high level of congestion of the city center throughout the day is more surprising. We can further investigate the temporal evolution of the spatial distribution of traffic. Of particular interest is the evolution of the cores from hour to hour, in particular between the morning and evening rush, as well as the evolution from year to year. Fig.~\ref{Evolution noyaux} displays the overlap between the cores of congested links, defined as links present in the giant component of congested links for more than $50\%$ of the days. The overlap is then calculated as the size of the intersection between the two cores, divided by the size of their union. These figures highlight the stability of the core during the morning or the evening rush hour. In addition, in 2014 there is an important overlap between congestion cores in the morning and in the afternoon. Even if there are differences between the morning and evening cores, this result points to the existence of `weak' links that are sensitive to demand and become easily congested, independently from the actual origins and destinations of the commuters. We note that the overlap between cores is weaker in 2018. Since we know that the average congestion on the network and the average size of the giant component have remained constant between 2014 and 2018 (figures \ref{Traffic measures Paris},\ref{size GC}), this indicates that the congestion must be more evenly spread on the network. In particular, it seems that the weak links, from which the giant component of congestion starts its progression, are not as clearly identified as in 2014. This is not necessarily good news, as it means that there is more randomness in the source of the giant component of congestion (any part of the network could potentially become congested randomly) and that there is no clear indication on where to put the focus for future road planning in order to mitigate the emergence of congestion. An interpretation for this difference between 2014 and 2018 could be the following. As we mentioned, an important highway along the Seine was closed in 2016. In 2014, most users travelling from West to East used this `Voie sur Berges', which carried high levels of traffic and therefore was prone to congestion. In 2018, the former users of this highway have been distributed on several alternative routes across the network. The city center in particular now operates at higher traffic density, which leads to more roads susceptible to trigger congestion.
%However, we note that this overlap between morning and evening hours is weaker in more recent years (2017, 2018), compared to 2014 or 2015. This suggests that different links are affected in the morning and in the evening signalling a lesser importance of weak points or bottlenecks that are systematically congested when the traffic demand is high {\bf relire et verifier}.
% Pas sûr qu'on puisse vraiment dire plus que ça ici

\subsection{Spatial contagion process}

Fig.~\ref{frequency distribution} yields other results than simply the study of the core of congested links. Indeed, we observe that the frequency at which links appear in the giant component follows a peculiar distribution that is essentially constant for the range $20\%-80\%$ (except for hours in the range 8pm-7am where we observe a strong decay). This is very different from what we would observe if the giant component
consists essentially of  the same links every day. In this case we would find a distribution with a high peak at large frequency and a smaller peak at low frequency. Here, in contrast we observe a whole range of frequency and we discuss here a simple toy model that allows us to understand this behavior. This model is based on the intuitive idea that congestion spreads from a link to its neighbor in a diffusive way.

Rather than building a reaction-diffusion process (like has been done by Bellocchi and Geroliminis \cite{Geroliminis diffusion}), we follow the results of \cite{Cogani real article, Jiang Havlin}) to build a much simpler model. In particular, it has been shown in \cite{Cogani real article} that the probability for two links to belong to the same functional cluster decays exponentially with distance, and that the probability of a link to become congested decays linearly with the distance to a congestion `seed'  \cite{Jiang Havlin}. We thus propose the following model: we assume that a single given link of the network can act as the seed for the giant component of congestion, and a link $i$ at distance $d_i$ from the seed link has a probability to belong to the giant component of congestion which reads as
\begin{align}
\label{equation p_i}
    p_i = \max(p_s(1-\mathrm{e}^{-d_i/L}), p_r)
\end{align}
where $p_s$ is the probability for congestion to emerge on the seed, and $p_r$ is a small residual probability, accounting for the fact that any link might be subject to a random incident causing its congestion. The probability to find a link (among a total of $N_L$ links) in $k$ of $N$ realizations of the congested network is then
\begin{align}
  \nonumber
  P(k,N)&=\frac{1}{N_L}\sum_i\binom{N}{k}p_i^k(1-p_i)^{N-k}\\
        &=\frac{\sum_d n(d)\binom{N}{k}p(d)^k(1-p(d))^{N-k}}{\sum_dn(d)}
          \label{eq:pkn}
\end{align}
where in the second line we replace the sum over links by a sum over the distance $d$ from the seed, weighted by the number $n(d)$ of links at distance $d$ from the seed. For a regular lattice in dimension $D$, $n(d)$ scales as $d^{D-1}$. However, for a real network the dependence between $n$ and $d$ might be different, for instance if the seed is placed in the periphery of the network. Moreover, in the case of emerging congestion on the network, it is likely that several links will act simultaneously as seeds for congestion. For these reasons, we expect in general to have $n(d) = d^\alpha$, where $\alpha \leq 1$ and can vary with the hour of the day. In addition, we expect $\alpha$ to be smaller for hours where the number of seeds in the network is larger such as during rush hours where the formation of congestion might happen on several links. In fact, the core of links of the giant component should be seen as the set of weakest links of the network on which jams are likely to appear spontaneously.

\begin{figure}[h!]
    \includegraphics[width=1.0\linewidth]{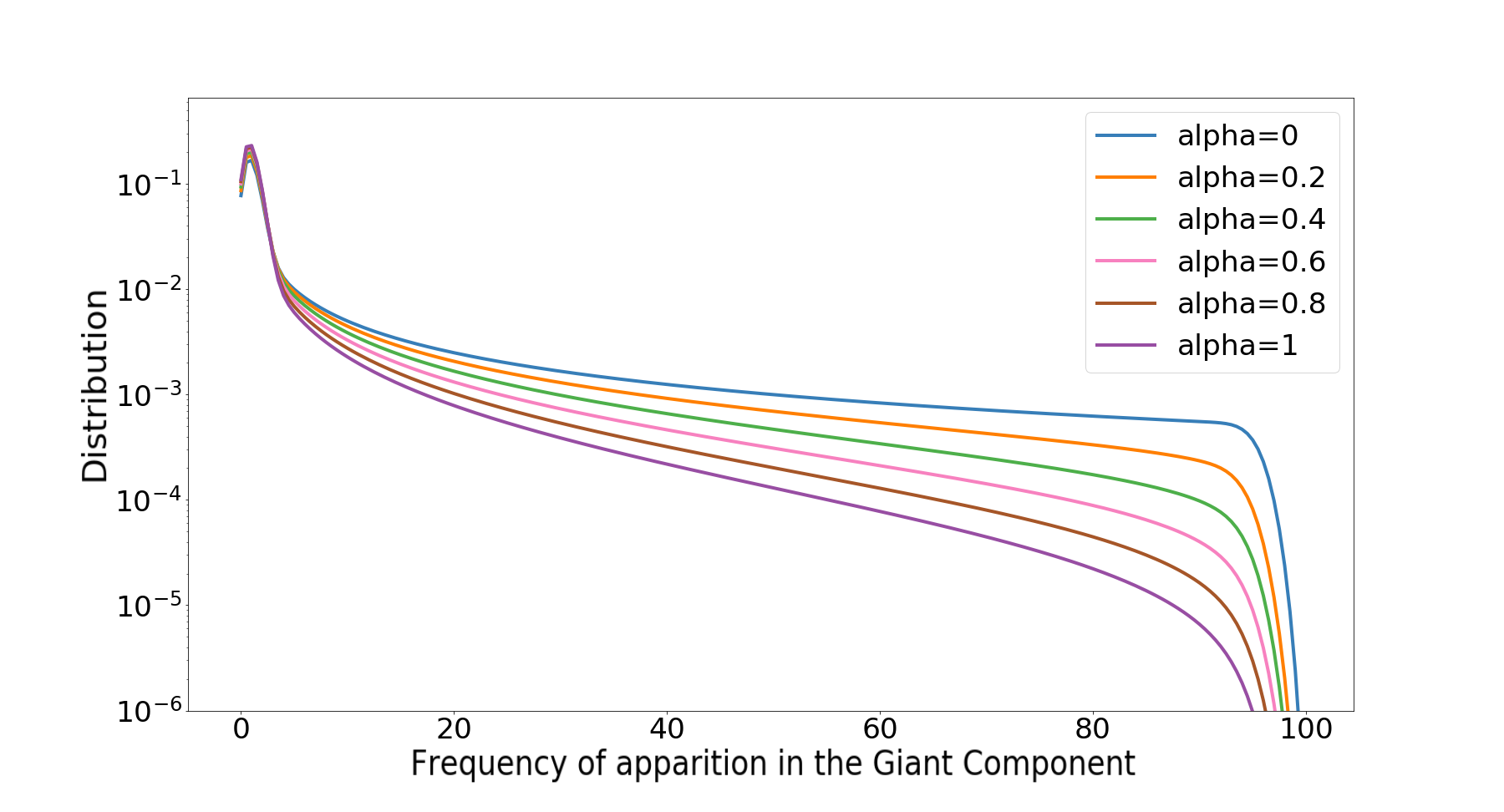}
    \caption{Numerical results for equation \ref{eq:pkn} for different distributions of links on the network $n(d)=d^\alpha$ (with $L=100m$, $p_s=0.95$ and $p_r=0.01$).}
    \label{test somme}
\end{figure}

We compute numerically the expression Eq.~\ref{eq:pkn} and show the result in Fig.~\ref{test somme} for different values of $\alpha$. Comparing this result to the empirical measure shown in Fig.~\ref{frequency distribution}, we find that for hours in the range 9am-5pm, we have $\alpha \approx 0$ which is consistent with the many congestion seeds observed during the day.  For hours in the range 6pm-8pm, we find a value closer to $\alpha\approx 1$ consistent with the fact that the congested core is much smaller in the evening (during night hours, the traffic demand is too low to lead to observe a growth of congestion from links acting as seeds). 
\vspace{-.8cm}

\section{Discussion}

In this work, we have shown that the delays occurring on the links of the network are correlated on a typical length which varies depending on the hour in the day. During rush hours, this correlation length becomes greater than the size of the network, indicating that the system undergoes a phase transition. The average occupancy rate or average density of cars on the network is a good measure of the traffic demand being put on the network, and acts as the control parameter of this phase transition. These results need to be confirmed for other cities and should be refined with datasets at a better resolution, in particular for computing critical exponents. We have also highlighted the unique role played by ring roads in urban traffic. This ubiquitous structure in large cities plays an essential role for commuters, but as we have shown here, has the side effect of contributing to quickly propagate congestion all over the network when it operates above its capacity. This knowledge is naturally critical for understanding urban traffic and especially for elaborating mitigation strategies. Finally, we investigated the spatial distribution of congested roads in Paris. We found further evidence to consider the jamming of the network as a process where congested links percolate throughout the network. Moreover, we identified the most vulnerable links, and proposed a model in which the congestion stems from those links and propagates to neighbouring links, confirming a reaction-diffusion type of picture for congestion spread. The observation of the time stability of the core of congested links led to the result that the vulnerable links vary depending on the hour. In particular, the traffic demand during the morning and the evening rush differ and the set of links prone to congestion are different in the morning and the evening. The knowledge of these weak links can be of interest for road planning and for a better theoretical understanding of traffic in a complex network of roads. 

%\FloatBarrier
%\vspace*{5cm}

\acknowledgments{} 
We thank the useful comments of anonymous referees. MB thanks J.-M. Luck for interesting references about the correlation function in statistical physics and bounds on the critical exponent $\eta$. Erwan Taillanter acknowledges the Institut de Physique Th\'eorique (IPhT, CEA Saclay)
for the funding of his thesis. 

%\newpage

%\addcontentsline{toc}{section}{References}
\bibliographystyle{apssamp}

\onecolumngrid
\newpage
\section*{Appendices}

\subsection*{Fundamental diagrams}

\begin{figure}[h!]
    \includegraphics[width=0.49\textwidth]{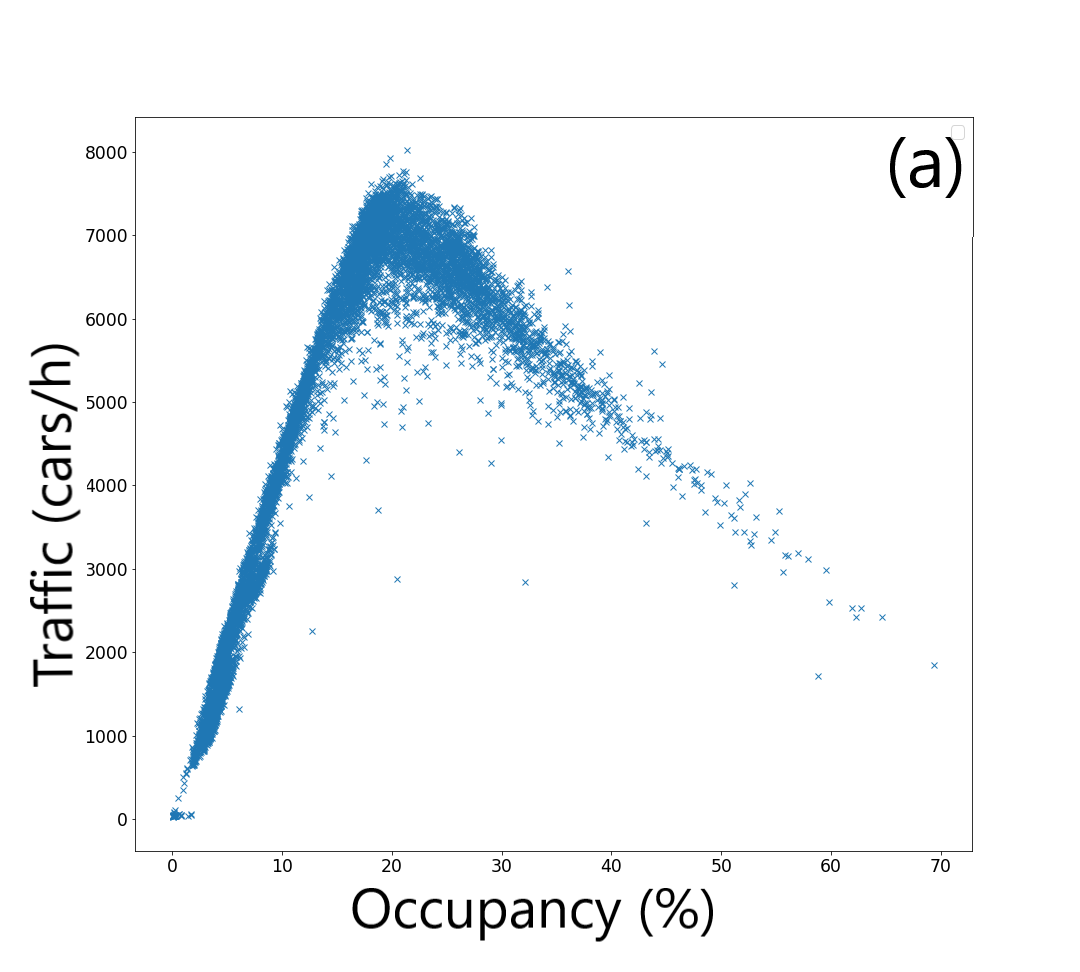}
    \includegraphics[width=0.49\textwidth]{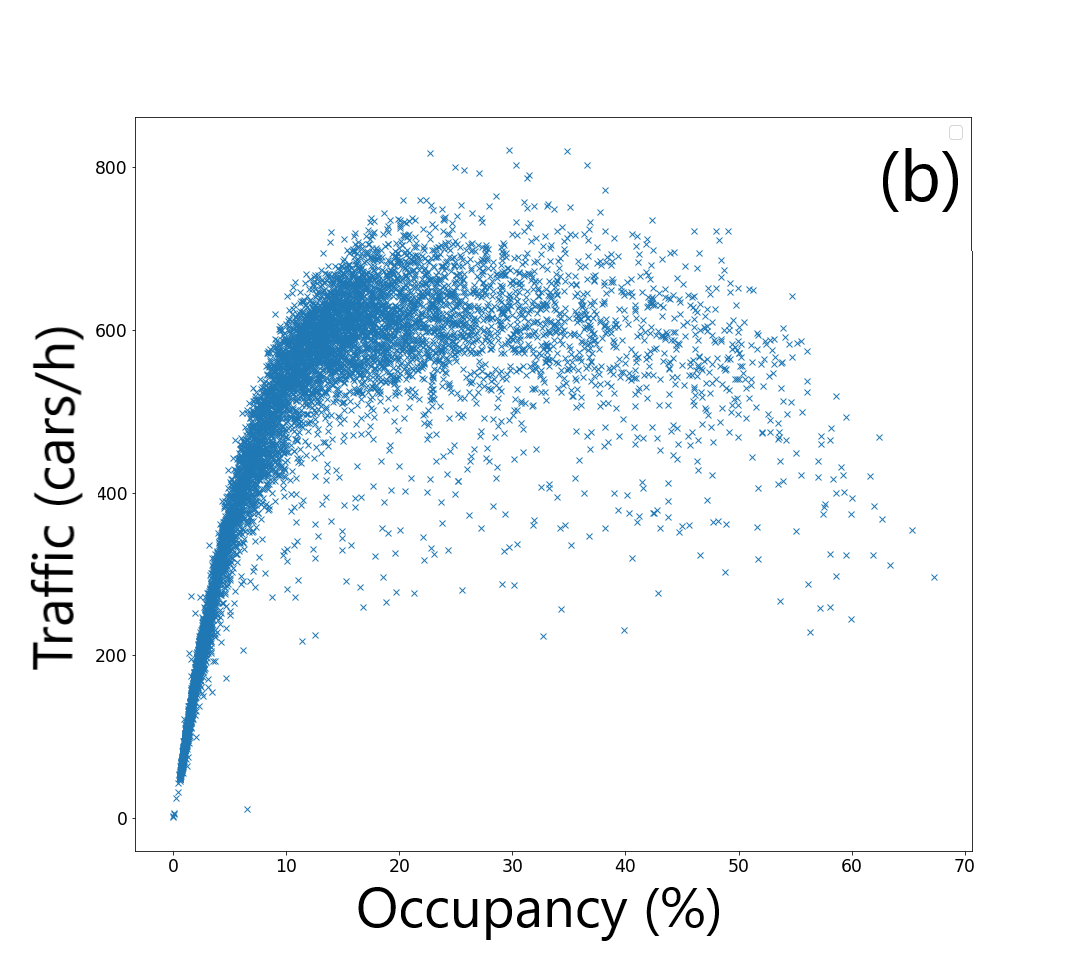}
    \caption{Fundamental diagrams for a segment of the Boulevard P\'eriph\'erique (a) and for a urban road - Avenue Jean Jaur\`es (b). Each point is an average measure of traffic and occupancy during one hour of the year 2014.}
    \label{Fundamental diagrams}
\end{figure}

The fundamental diagram is the cornerstone of standard traffic theory. Assuming an infinite, one lane road, one has the following situation. At low car density, each car drives at the speed limit for this road, regardless of other cars. For small densities, one thus has $v(k) = \textrm{constant} = v_{max}$ and $q = v(k)*k$ proportional to $k$. At higher car densities, however, cars will be closer to one another. Drivers tend to reduce their speed as the distance to the preceding car decreases. For a given density, known as \textit{jamming density} and corresponding to circa 1m between cars, the traffic comes to a complete stop: the road is jammed and $q(k_{jam}) = 0$. Between $k=0$ and $k=k_{jam}$, the flux of vehicles reaches a maximum and then decreases. On one lane roads or highways, $q(k)$ is known to be well defined, and has a typically triangular shape. On a urban network, things are however more complex. Indeed, the relation between $q$ and $k$ is also influenced by many other factors, such as traffic lights or the traffic state of roads met at intersections. Figure \ref{Fundamental diagrams} displays the difference between the fundamental diagram of the ring road (a textbook fundamental diagram for highways) and a fundamental diagram observed for a link of the urban network. As one can see, the latter is not as well defined, with points spread broadly around the average curve. Besides, the form of the urban diagram is not triangular but rather plateauing at high values of $q$, indicating that the traffic is not defined solely by the density, but indeed by other external factors.

However, for any road, the relation $v_{max} = \frac{dq}{dk}(0)$ remains true, which is the most valuable information we extract from these diagrams in this work.

\subsection*{Correlation function fit}

We give examples of correlation functions and the corresponding fit. As mentioned, distances shorter than a minute of travel time correspond to roads separated by just one junction, for which the correlation is determined mostly by traffic light timing. The largest distance between roads ($\approx 5000$ s) corresponds to a very small amount of pairs. We thus perform our fit on the range $[60,3000]$ seconds of travel time. We find an exponent $\eta \approx 0.4$.

\begin{figure}[h!]
    \includegraphics[width=0.49\textwidth]{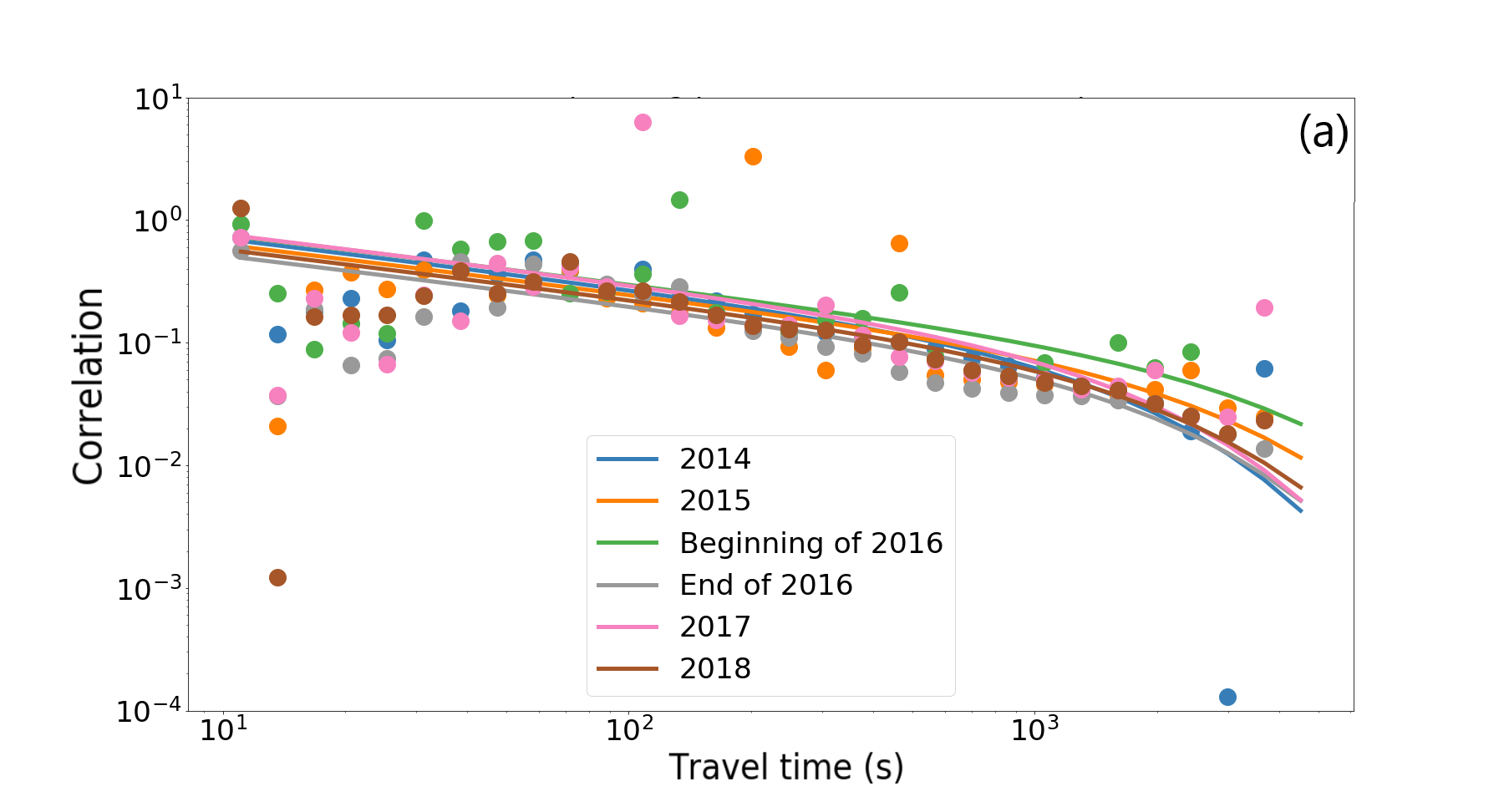}
    \includegraphics[width=.49\textwidth]{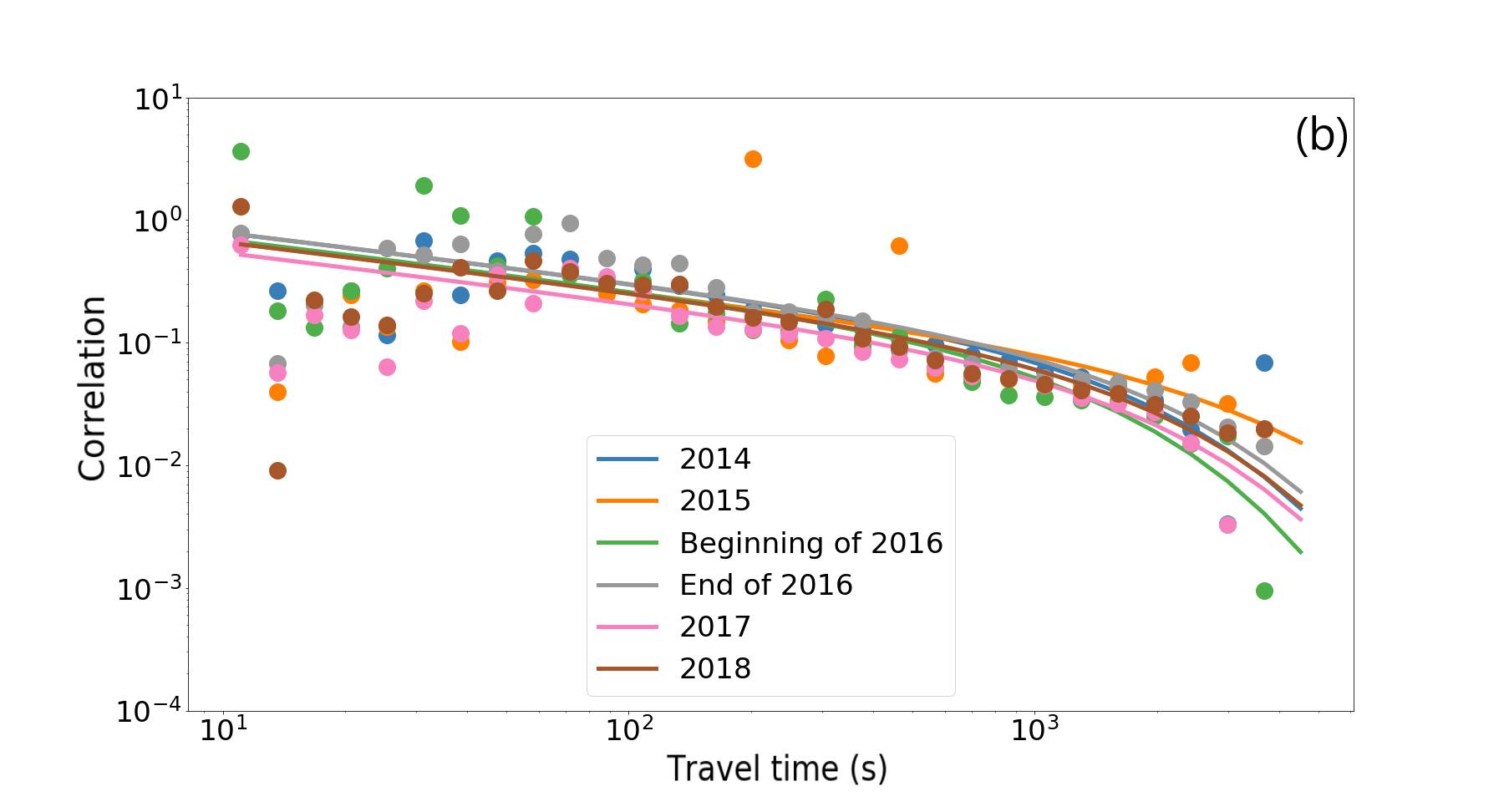}\\
    \includegraphics[width=.49\textwidth]{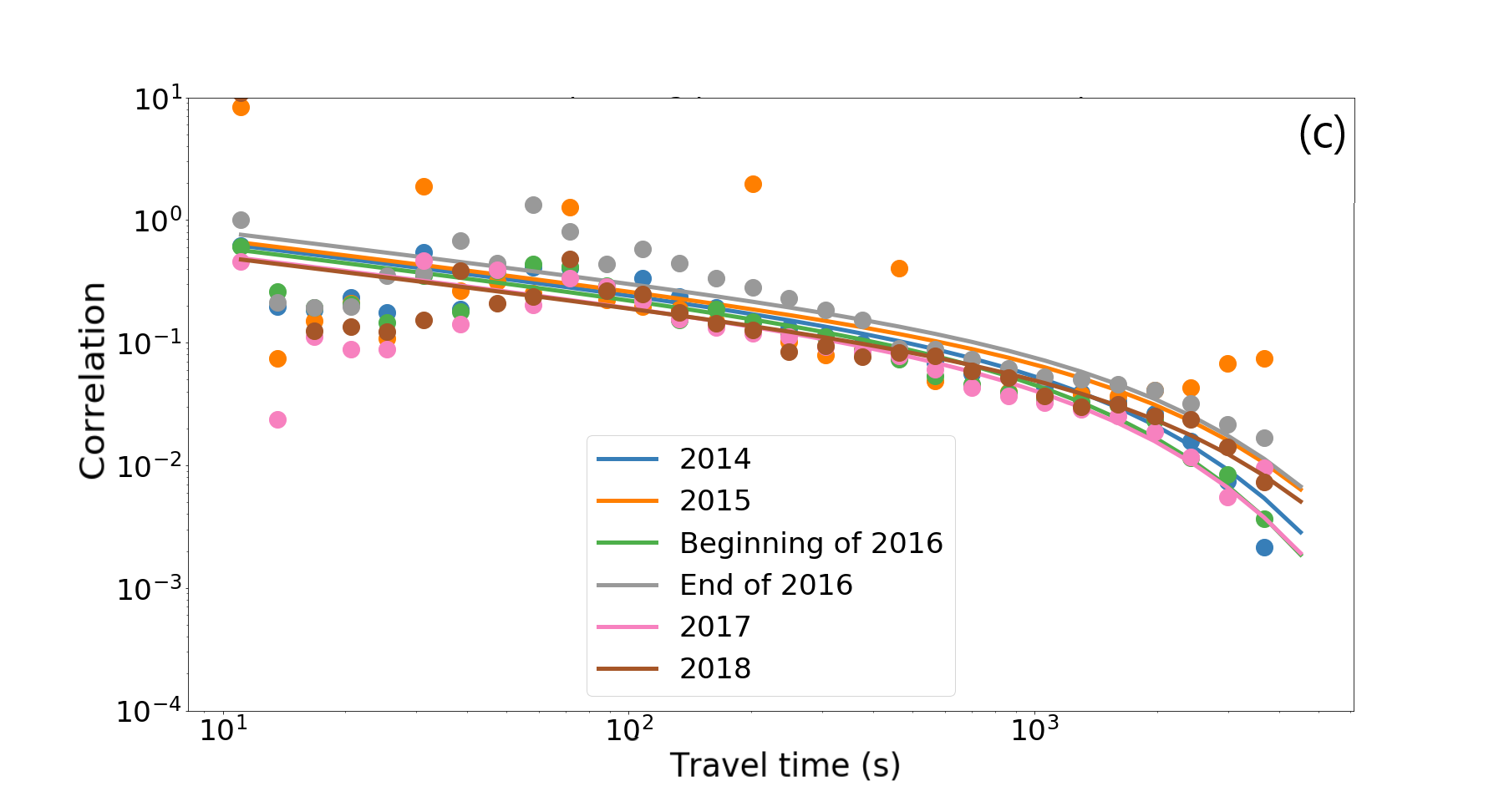}
    \includegraphics[width=.49\textwidth]{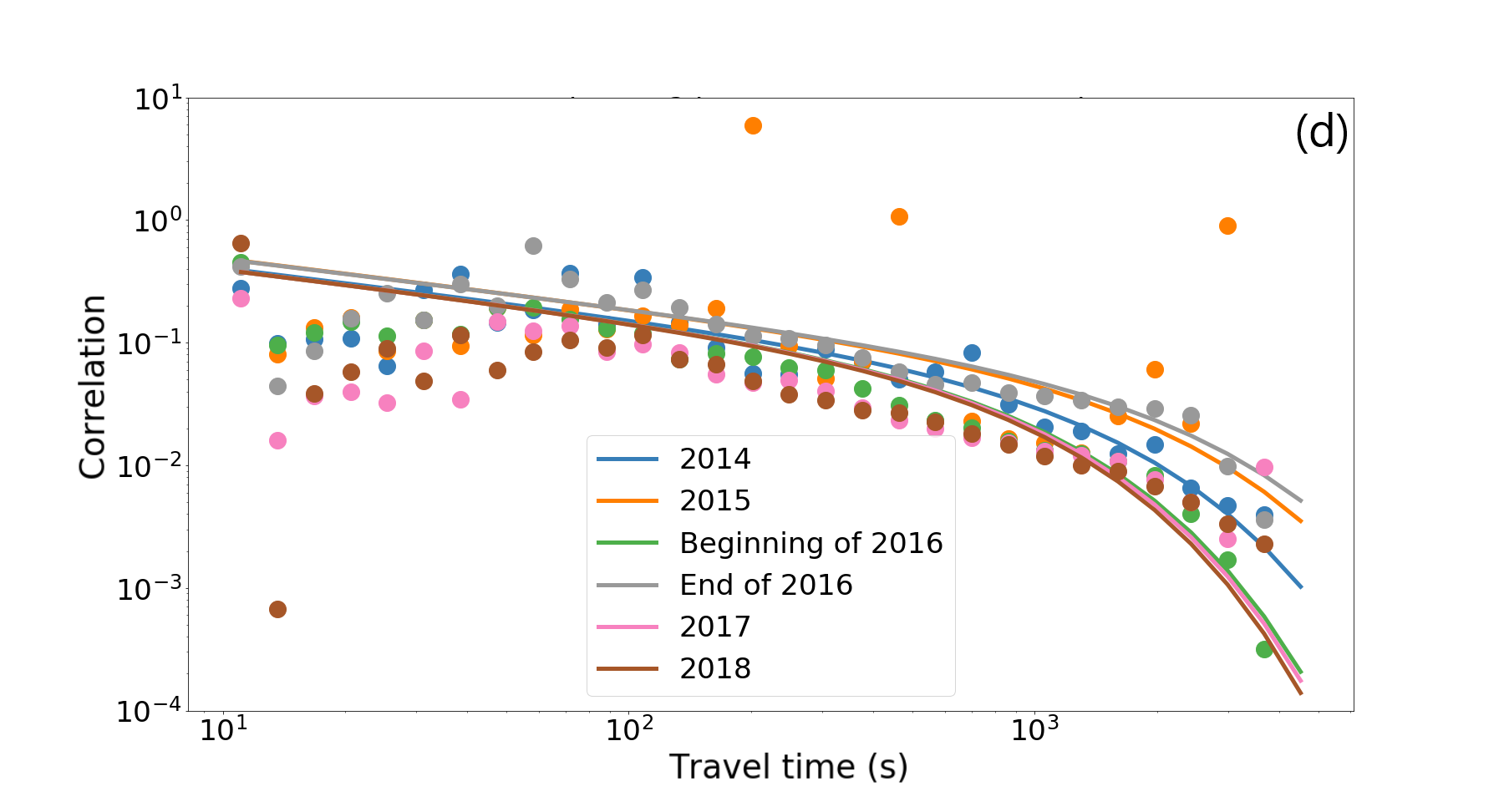}
    \caption{Illustration of the correlation functions and the corresponding fit for hours 18 to 21 respectively. The distance is expressed in terms of travel time along the network. The exponent of the power law decay is $\eta \approx 0.4$}
    \label{Fits correl}
\end{figure}

\subsection*{Travel time versus Euclidean distance}
For most distances, the travel time scales as $d^{\beta}$ with $\beta \approx 0.4$, while at larger distances, the travel time becomes independent of the distance. This effect is of course an artefact of the geometry of our network. Indeed, we only consider Paris and its ring road. The points furthest away are diametrically opposed points on the ring road, between which the travel can be made at high speeds. Points slightly closer than that from each other will most likely need a fair bit of travel inside the urban network, explaining why it doesn't translate into shorter travel times. If we had considered a larger network, including roads offside of the 'Boulevard Périphérique', we would most likely not see this effect.

\begin{figure}[h!]
    \centering
    \includegraphics[width=.7\textwidth]{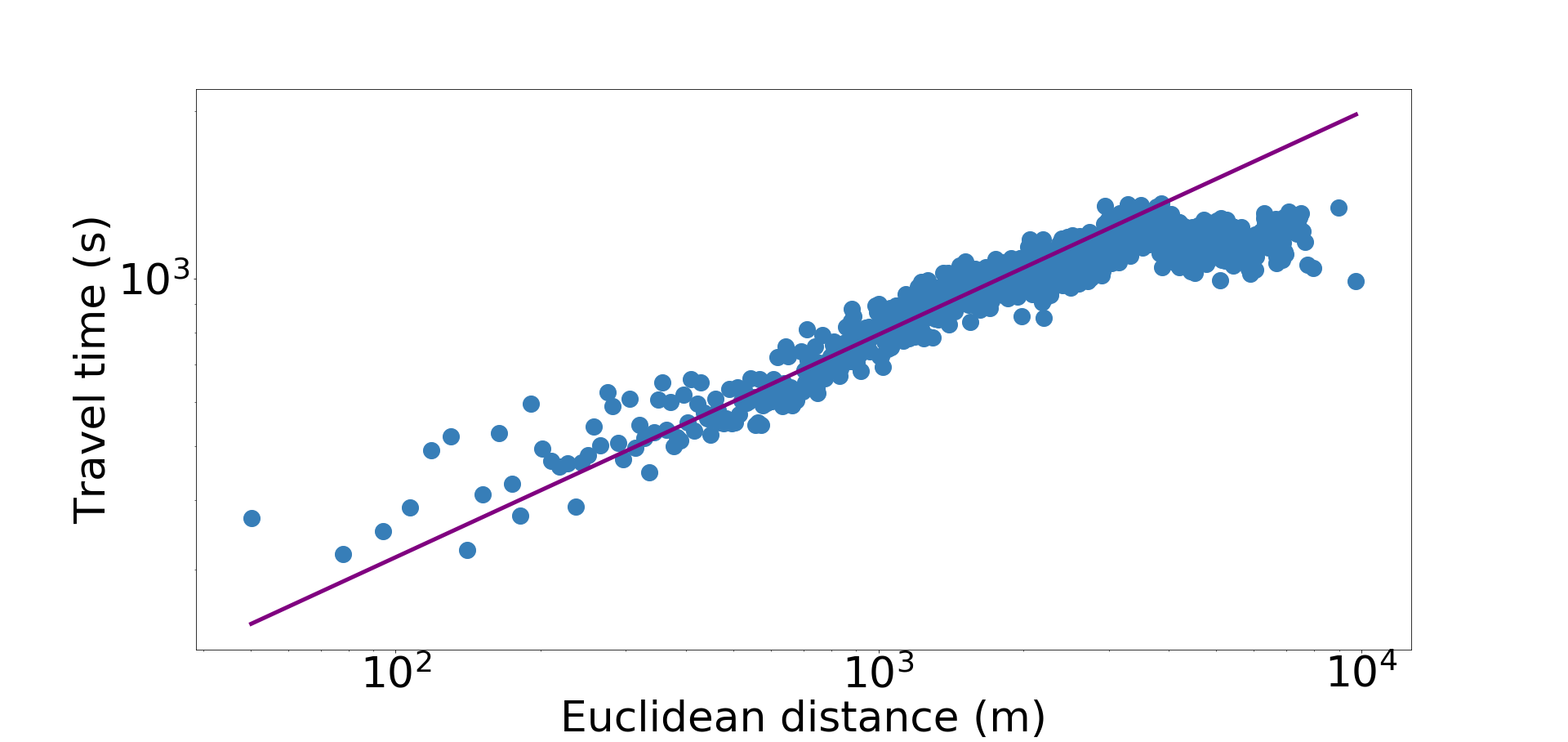}
    \caption{Relation between travel time and euclidean distance in the case of the Parisian network. Corresponding powerlaw fit of exponent $\beta \approx 0.4$}
    \label{travel time vs distance}
\end{figure}

\subsection*{Congestion cores}

Here we give example of the frequency of presence of the links of the network in the giant component of congestion for different hours of the year 2014. We see that the overall congestion is not the the only factor determining the links which are likely to collapse, but the origins and destinations of commuters seem to play a major role as well. Indeed, the overall congestion is lower at noon, yet the center of Paris is very prone to congestion and a significant part of it is almost always in the giant component. During the evening rush, on the other hand, the average congestion is high. However, it seems that it is quite evenly distributed on the network, and the Giant component of congestion is variable from day to day, with no link systematically present in it.

\begin{figure}[h!]
    \includegraphics[width=0.49\textwidth]{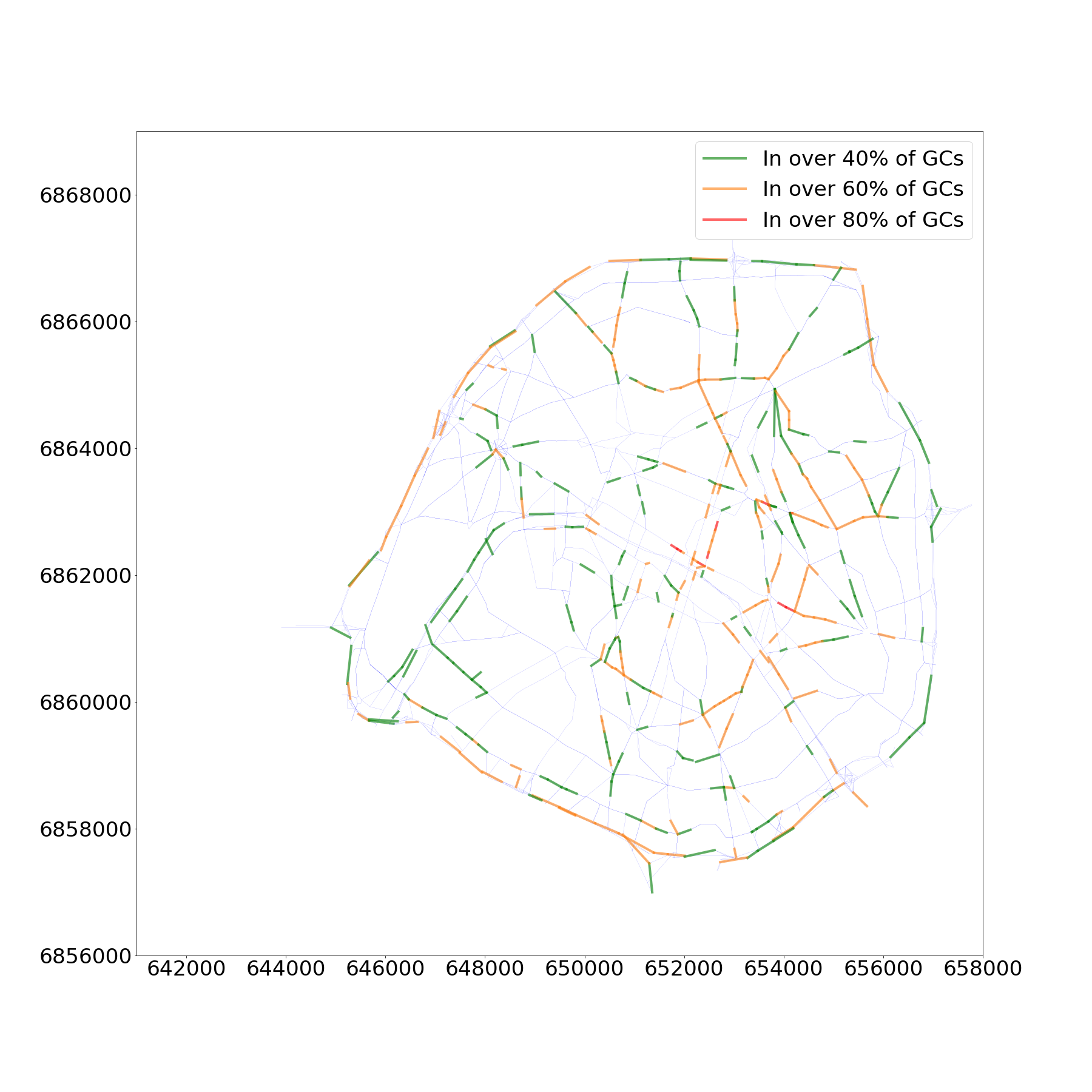}
    \includegraphics[width=0.49\textwidth]{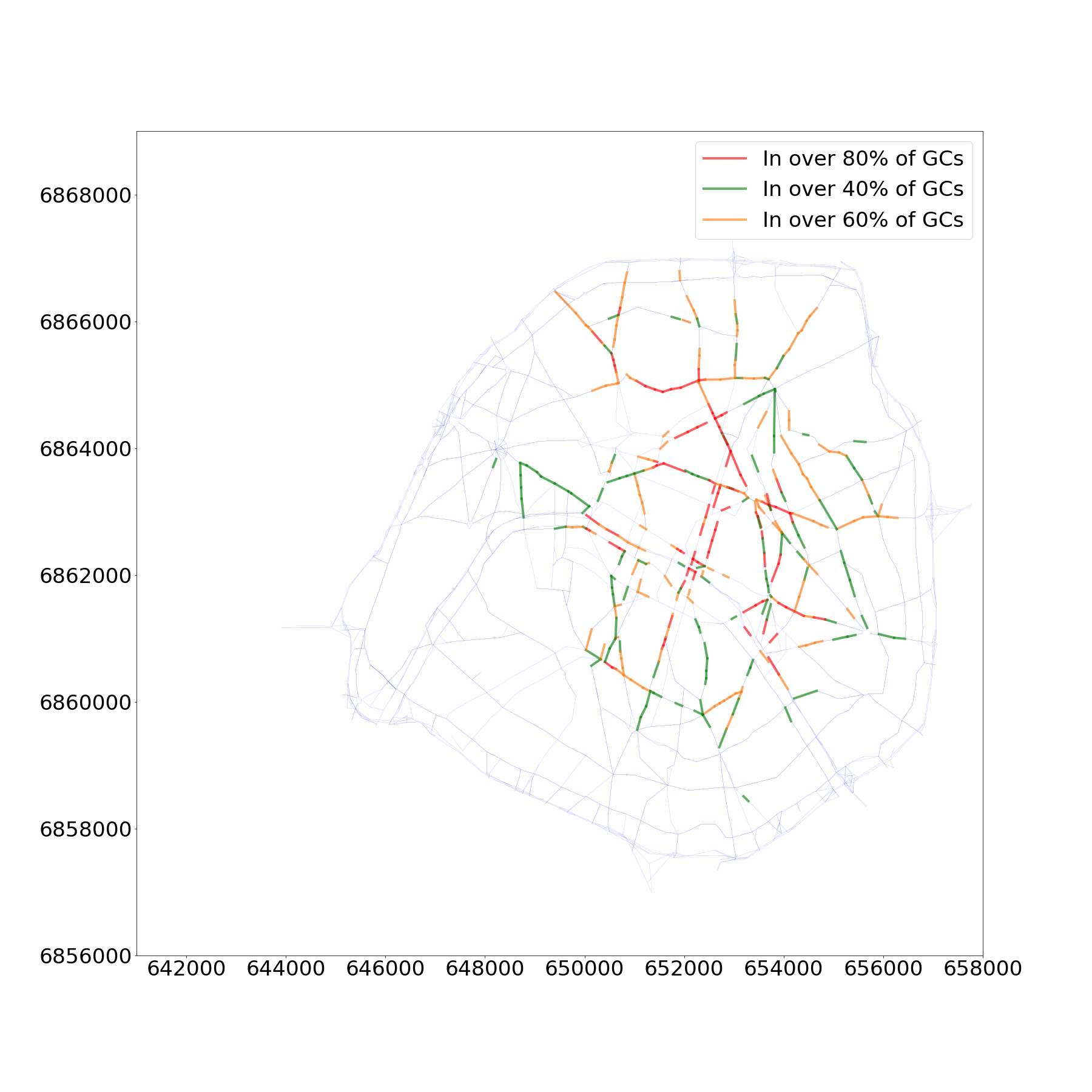}
    \caption{Heatmap of the frequency of presence of the links in the Giant Component of congested links, highlighting the core of links likely to be the seeds for congestion, at 9am and 12am respectively in 2014}
    \label{Noyaux1}
\end{figure}

\begin{figure}[h!]
    \includegraphics[width=0.49\textwidth]{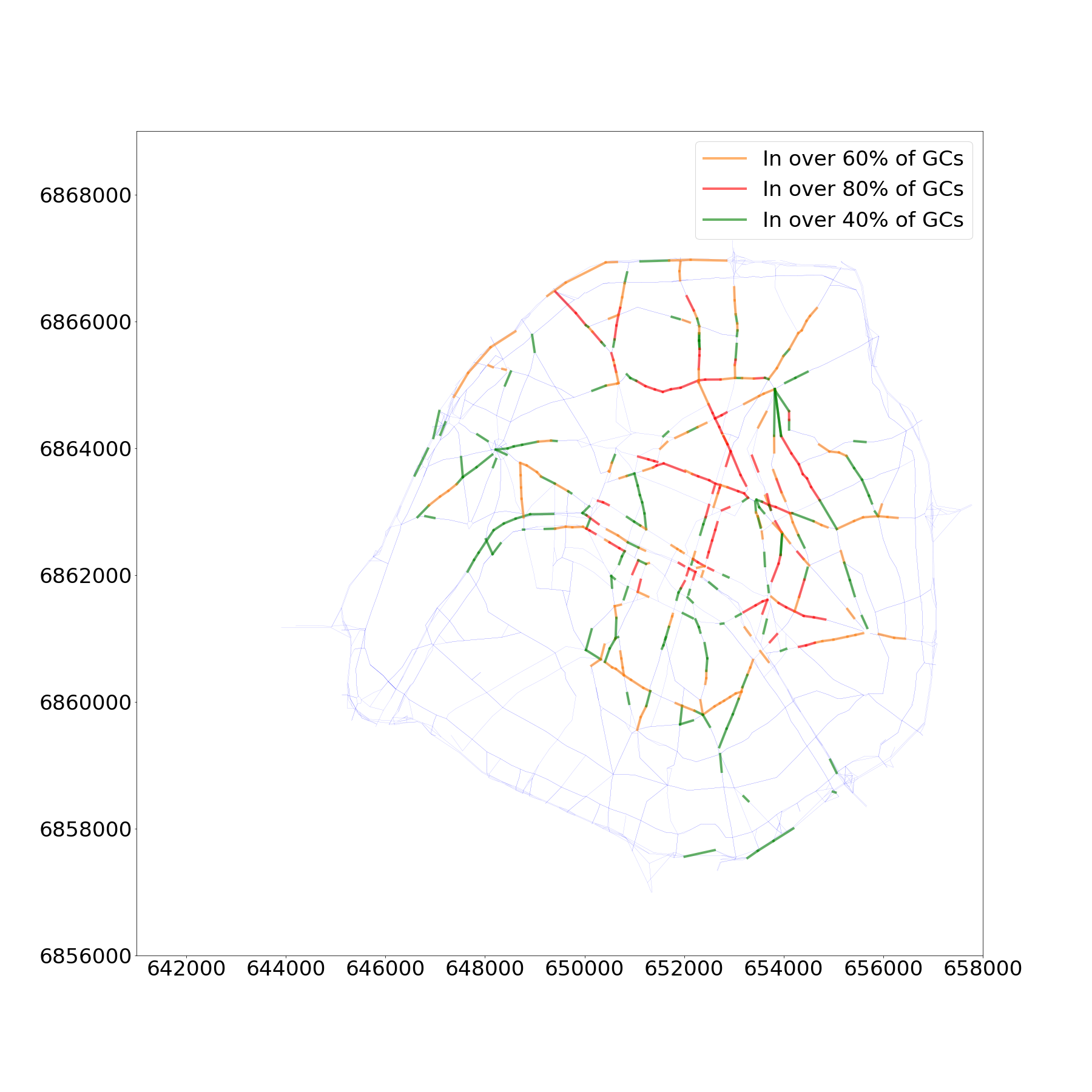}
    \includegraphics[width=0.49\textwidth]{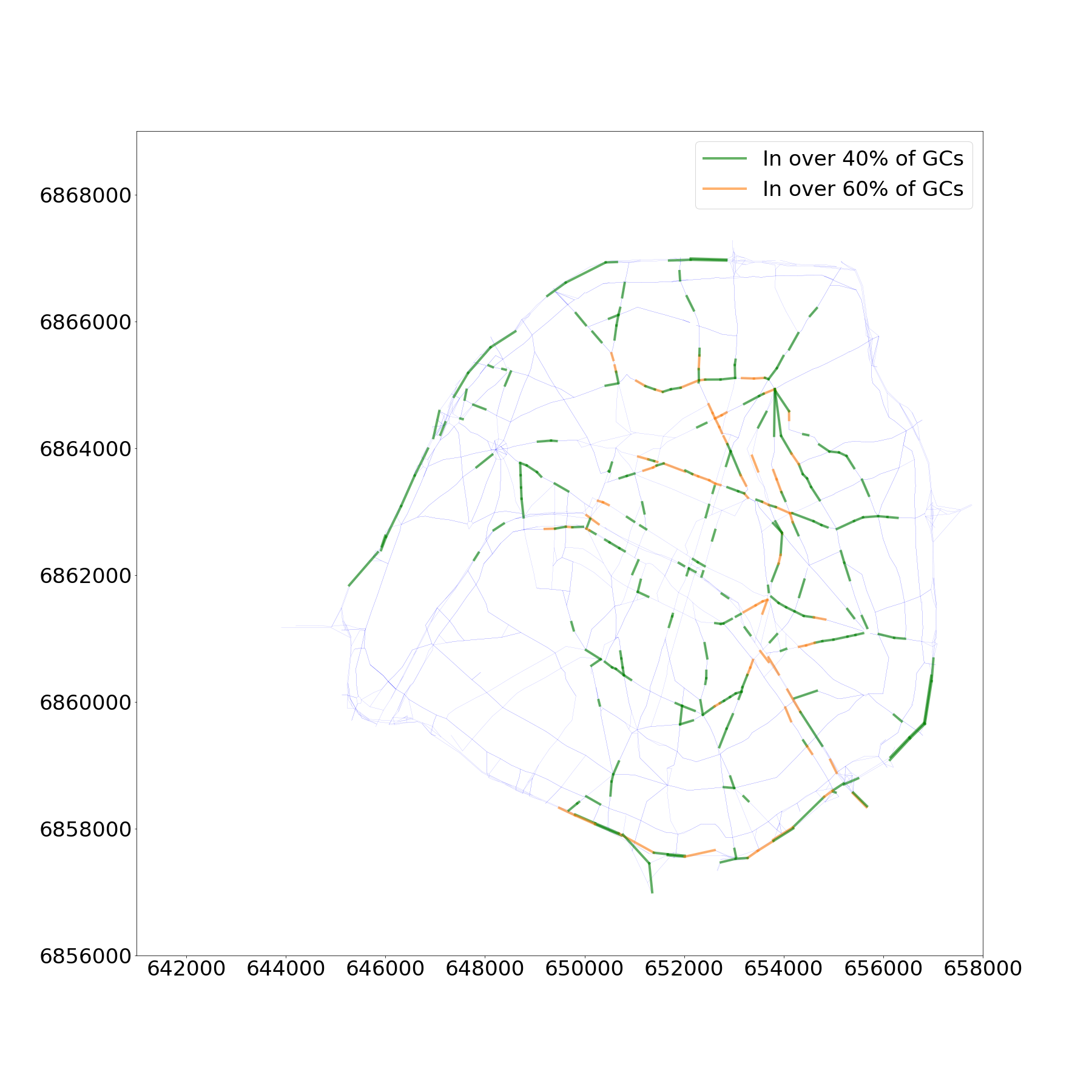}
    \caption{Heatmap of the frequency of presence of the links in the Giant Component of congested links, highlighting the core of links likely to be the seeds for congestion, at 4pm and 7pm respectively in 2014}
    \label{Noyaux2}
\end{figure}

\end{document}